\documentclass[flushrt]{emulateapj}

\usepackage{subfigure} 
\usepackage{epsfig}
\newcommand{\npar}{\par \vspace{2.3ex plus 0.3ex minus 0.3ex}}
\usepackage{lscape}
\usepackage{longtable}

\begin{document}
\title{A Magnified View of the Kinematics and Morphology of RCSGA 032727-132609: Zooming in on a Merger at $z=1.7$} 

\author{Eva Wuyts\altaffilmark{1}, Jane R. Rigby\altaffilmark{2}, Michael D. Gladders\altaffilmark{3,4}, Keren Sharon\altaffilmark{5}}
\altaffiltext{1}{Max-Planck-Institut f\"{u}r extraterrestrische Physik, Postfach 1312, Giessenbachstr., D-85741 Garching, Germany}
\altaffiltext{2}{Observational Cosmology Lab, NASA Goddard Space Flight Center, Greenbelt, MD 20771}  
\altaffiltext{3}{Department of Astronomy and Astrophysics, University of Chicago, 5640 S. Ellis Av., Chicago, IL 60637}
\altaffiltext{4}{Kavli Institute for Cosmological Physics, University of Chicago, 5640 South Ellis Avenue, Chicago, IL 60637}
\altaffiltext{5}{Department of Astronomy and Astrophysics, University of Michigan}

\begin{abstract}
We present a detailed analysis of multi-wavelength HST/WFC3 imaging and Keck/OSIRIS near-IR AO-assisted integral field spectroscopy for a highly magnified lensed galaxy at $z=1.70$. This young starburst is representative of UV-selected star-forming galaxies (SFG) at $z\sim2$ and contains multiple individual star-forming regions. Due to the lensing magnification, we can resolve spatial scales down to $100$~pc in the source plane of the galaxy. The velocity field shows disturbed kinematics suggestive of an ongoing interaction, and there is a clear signature of a tidal tail. We constrain the age, reddening, SFR and stellar mass of the star-forming clumps from SED modelling of the WFC3 photometry and measure their H$\alpha$ luminosity, metallicity and outflow properties from the OSIRIS data. 
With strong star formation driven outflows in four clumps, RCSGA0327 is the first high redshift SFG at stellar mass $<10^{10}$~M$_\odot$ with spatially resolved stellar winds. We compare the H$\alpha$ luminosities, sizes and dispersions of the star-forming regions to other high-z clumps as well as local giant H~II regions and find no evidence for increased clump star formation surface densities in interacting systems, unlike in the local Universe. 
Spatially resolved SED modelling unveils an established stellar population at the location of the largest clump and a second mass concentration near the edge of the system which is not detected in H$\alpha$ emission. This suggests a picture of an equal-mass mixed major merger, which has not triggered a new burst of star formation or caused a tidal tail in the gas-poor component. 
\subjectheadings{galaxies: high-redshift, strong gravitational lensing, galaxies: kinematics and dynamics, galaxies: structure} 
\end{abstract}

\section{Introduction}
\label{sec:intro}
High redshift star-forming galaxies (SFG) show increasingly irregular and clumpy morphologies compared to the local Universe. This was first observed at rest-frame UV wavelengths with the Hubble Space Telescope (HST) \citep{Griffiths1994,vandenbergh1996,Elmegreen2007}, and has since been confirmed in rest-frame optical light \citep{Elmegreen2009,Forster2011,Guo2012}, H$\alpha$ emission \citep{Swinbank2009,Jones2010, Genzel2011, Livermore2012, Wisnioski2012}, and sub-millimeter emission \citep{Swinbank2010}. Locally, such irregular morphologies are associated with mergers, but a range of recent observational findings point toward this not being the common explanation at high redshift. The existence of a tight correlation between star formation rate (SFR) and stellar mass out to at least $z\sim2.5$, referred to as the main-sequence of star formation, favors continuous star formation activity over a series of rapid, luminous merger-driven bursts \citep{Noeske2007,Elbaz2007,Daddi2007,Wuyts2011}. Outliers of the main-sequence account for only 10\% of the cosmic star formation density at $z\sim2$ \citep{Rodighiero2011}. Counts of both galaxy pairs and disturbed morphologies \citep{Conselice2009,Lotz2011,Kaviraj2013}, as well as studies of gas-phase kinematics \citep{Forster2009,Epinat2012}, limit the $z\sim2$ merger rate to up to 30\%. Based on the high gas fractions of 30-80\% \citep{Daddi2010,Tacconi2010,Tacconi2012} and high velocity dispersions of 50-100~km/s \citep{Forster2009, Law2009, Wisnioski2011} of $z\sim2$ SFGs, an alternative picture has developed where luminous kpc-sized clumps are formed through gravitational instabilities in a dynamically unstable, gas-rich, turbulent disk \citep{Noguchi1999, Immeli2004a, Immeli2004b, Bournaud2007, Dekel2009, Genel2012}. If they can survive long enough, these clumps are expected to migrate towards the center of the galaxy due to dynamical friction and coalesce into a young bulge on timescales of $\sim0.5$~Gyr. In this scenario the galaxy's gas reservoir is continuously replenished by the accretion of gas from the halo through minor mergers and cold flows, to sustain the observed large gas fractions and strong star formation activity \citep{Keres2005, Keres2009, Bournaud2009, Dekel2009}. These observational and theoretical results favour internal secular evolution over major mergers to explain the high star formation density and clumpy morphology of SFGs at $z\sim2$. However, merging is still thought to play an important role in the cosmological mass assembly of galaxies, the quenching of star formation and the morphological transitions of galaxies from late to early-type (e.g. \citealt{Springel2005, Naab2003, Naab2007, Guo2008, Hopkins2010}).
\npar
Current observational studies of the kinematics of high-z SFGs and the properties of their individual star-forming regions are limited by the available spatial resolution. At FWHM $\sim 0\farcs1$, which corresponds to roughly 1~kpc at $z=2$, both HST imaging and adaptive optics (AO) assisted integral field spectroscopy (IFS) at 8-10~m class telescopes barely resolve the largest clumps. Beam-smearing of the velocity field can compromise the kinematic classification of merger signatures as well as the analysis of turbulence, shocks and outflows from the velocity dispersion map (e.g. \citealt{Kronberger2007, Davies2011}).

Strong gravitational lensing of high redshift background galaxies by foreground galaxy clusters can increase their apparent size by an order of magnitude or more. IFS studies of lensed galaxies have already provided detailed views of galaxy kinematics up to $z=4.9$ \citep{Nesvadba2006, Swinbank2009, Jones2010, Jones2013, Yuan2011, Yuan2012, Shirazi2013}, confirming the common occurrence of rotating disks at high redshift. The improved spatial resolution has allowed these studies to specifically address shocks and outflows \citep{Yuan2012}, metallicity gradients \citep{Yuan2011, Jones2013}, and the properties of star-forming clumps \citep{Swinbank2009, Jones2010}. In this paper we present a detailed analysis of HST/WFC3 optical/near-IR imaging and AO-assisted Keck/OSIRIS IFS data for the brightest distant lensed galaxy currently known in the Universe, RCSGA 032727-132609 at $z=1.7$ \citep{me2010}. This combined dataset probes the galaxy kinematics as well as the morphology of both the ongoing star formation and the established stellar population at spatial resolutions down to $\sim100$pc in the galaxy's source plane. 
\npar
The paper is organised as follows. \S\ref{sec:data} describes the main Keck/OSIRIS integral field spectroscopy observations and data reduction as well as supporting datasets from HST/WFC3, Keck/NIRSPEC and Magellan/FIRE. The kinematic analysis of the IFS data is presented in \S\ref{sec:kin} and \S\ref{sec:clumps} reports on the physical properties of the individual star-forming clumps that can be identified in both datasets. The radial variations of these properties across the galaxy are studied in \S\ref{subsec:radial} and the clumps are compared to scaling relations of local H~II regions in \S\ref{subsec:scaling}. We present spatially resolved SED modelling of the system in \S\ref{sec:spatialsed}. \S\ref{sec:disc} summarises the observational results and discusses what we can infer for the physics governing the morphology and kinematics of the system. Throughout this work, we adopt a flat cosmology with $\Omega_M = 0.3$ and H$_0 = 70$\,km\,s$^{-1}$\,Mpc$^{-1}$. All magnitudes are quoted in the AB system. 

\section{Observations and Data Reduction}
\label{sec:data}
RCSGA 032727-132609, hereafter RCSGA0327, at $z=1.703$ is the brightest and most obvious strong lensing system found in the Second Red-Sequence Cluster Survey (RCS2; \citealt{Gilbank2011}). Its discovery, preliminary lensing analysis and global galaxy properties are presented in \cite{me2010,me2012a}. The system consists of a counter-image, and a giant arc which extends over 38\arcsec\ on the sky and is made up of three merged images of the background source. Its intrinsic stellar mass of $6.3\pm0.7 \times 10^9$~M$_\odot$ and SFR of 30-50~M$_\odot$~yr$^{-1}$ \citep{me2012a} translate into a specific star formation rate $\log(\mathrm{sSFR}/\mathrm{yr}^{-1})=-8.3$, which lies a factor of 3 above the main sequence of star formation at $z\sim2$ \citep{Daddi2007}. This paper focuses on the joint analysis of HST/WFC3 imaging and Keck/OSIRIS integral field spectroscopy observations of the arc.  To support the analysis, we fold in additional long-slit near-infrared spectroscopy from the NIRSPEC instrument on Keck, and the FIRE instrument on Magellan.

\begin{figure*}
\centering
\includegraphics[width=\textwidth]{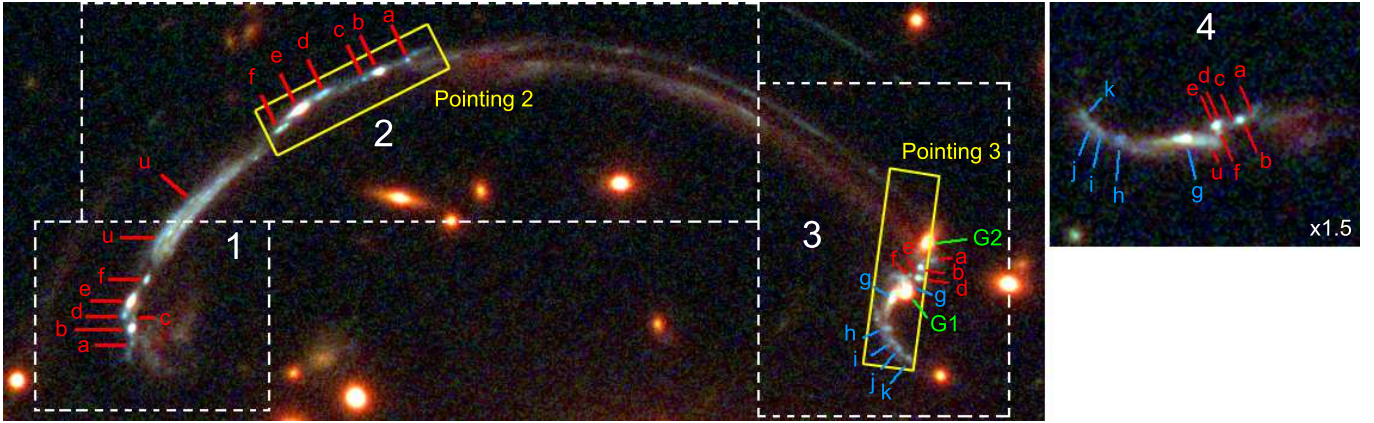}
\caption{Identification of substructure in the giant arc and counter-image of RCSGA0327 from HST/WFC3 imaging; North is up and East is left. The color rendition is composed of F160W+F125W+F098M (red); F814W+F606W (green); and F390W (blue), to highlight color gradients in the arc. The left panel extends over 35\arcsec\ $\times$ 15\arcsec. The dashed lines approximately enclose the parts of the arc that compose each of the three merged images, indicated by numbers 1-3; image 4 corresponds to the counter-image, which is a relatively undistorted image of the source-plane galaxy. Individual star-forming regions are labeled with letters $a$ through $k$; the two cluster galaxies that fall on top of image 3 are labeled $G1$ and $G2$ in green. Due to the location of the source galaxy with respect to the lensing caustic, the western side (clumps $a$-$f$; red labels) appears in all of the four images, while the brightest knot of the source (labeled $g$) and the eastern ``arm'' (clumps $h$-$k$; blue labels) only appear in images 3 and 4. In image 3, the lensing perturbation by cluster galaxy $G1$ results in another instance of clump $g$. The two OSIRIS pointings with a $1\farcs6 \times 6\farcs4$ field of view and position angles of 297\arcdeg\ and 352\arcdeg\ East of North are overlaid in yellow. Figure adapted from \cite{Sharon2012}. \label{fig:implane}}
\end{figure*}

\subsection{HST/WFC3 Imaging}
\label{subsec:data-hst}
RCSGA0327 was imaged with the Wide Field Camera 3 (WFC3) on HST under GO program 12267 (PI: Rigby). In four orbits, the source was observed with one narrow-band filter targeting H$\beta$ (F132N) and six broad-band filters (F390W-F606W-F814W in the UVIS channel, F098M-F125W-F160W in the IR channel). The imaging strategy consisted of four sub-pixel dither positions in each filter to reconstruct the PSF, reject cosmic rays and compensate for the chip-gap. Individual frames were processed with the standard WFC3 calibration pipeline, and combined using the Multidrizzle routine \citep{Koekemoer2002}. Based on the HST data, \cite{Sharon2012} have matched the substructure in the four different images of the source galaxy to create a robust and well-constrained lens model with magnification uncertainties $\le$10\%. We adopt the naming convention introduced there and shown in Figure~\ref{fig:implane} to identify the star-forming clumps. Source-plane reconstructions of the galaxy are created by ray-tracing the image pixels through deflection maps generated from the lens model. 

\subsection{Keck/OSIRIS observations}
\label{subsec:data-osiris}
We observed RCSGA0327 with the OH-Suppressing Infrared Imaging Spectrograph (OSIRIS; \citealt{Larkin2006}) on the Keck~II telescope on two half nights of 2011, October 17-18 UT. Conditions were good with seeing measured at 0.7-1.2\arcsec. OSIRIS is an integral field spectrograph with a lenslet array to provide simultaneous near-IR spectra at spectral resolution $R\sim3600$ for up to 3000~pixels in a rectangular field of view up to $4\farcs8 \times 6\farcs4$.
Due to a technical difficulty with OSIRIS at the time of observations, only the broadband filters were available. We targeted the H$\alpha$ emission line with the Hbb filter in the $0\farcs1$ pixel$^{-1}$ scale, which limits the field of view to $1\farcs6 \times 6\farcs4$. To correct for atmospheric distortion, the laser-guide star adaptive optics system (LSGAO; \citealt{Wizinowich2006,vanDam2006}) was applied with a tip-tilt star of $R=17.0$ at a distance of $\sim40$\arcsec\ from the giant arc. This delivers a Strehl ratio of $\sim0.2$. Unfortunately, no suitable tip-tilt star is available for LSGAO observations of the counter-image. Due to the limited field of view of the spectrograph and the large size of the giant arc, we were not able to observe its full extent. Figure~\ref{fig:implane} shows the two pointings we selected in the image plane; we will refer to them as pointing 2 and 3 since they target parts of the arc that correspond to images 2 and 3 of the source galaxy. Pointing 2 at a position angle of 297\arcdeg\ East of North corresponds to one of the most highly magnified regions of the giant arc \citep{Sharon2012}, where we can take maximal advantage of the increased brightness and spatial resolution to study clumps $a$ through $f$. Pointing 3 at a position angle of 352\arcdeg\ East of North was chosen because it covers the full extent of the source. The early-type cluster galaxies $G1$ and $G2$ that fall on top of the arc in this pointing are not expected to show any emission lines, so they should not contaminate the H$\alpha$ flux.

The observations started with short 30\,s exposures of the tip-tilt star to center the pointing. These exposures are also used to calculate the point spread function (PSF): 2D Gaussian fits to the tip-tilt exposures yield a FWHM resolution of $\sim0.15$\arcsec. From the tip-tilt star, we applied a blind offset to acquire each of the two pointings. Individual science exposures have an integration time of 900\,s and are dithered by up to $0\farcs5$ around the base position to remove bad pixels and cosmic rays. Off-source sky-frames were necessary because the source fills most of the narrow $1\farcs6 \times 6\farcs4$ field of view. We achieved a total on-source integration time of 1.5~hours for pointing 2 and 3.5~hours for pointing 3. 

Data reduction was carried out with the OSIRIS data reduction pipeline (version 2.3)\footnotemark[1], which removes crosstalk, detector glitches, and cosmic rays and performs a scaled sky subtraction based on \cite{Davies2007}. Individual datacubes are mosaicked using a 3$\sigma$ clipping average and the final datacube is flux calibrated based on the 2MASS $H$-band magnitude of the tip-tilt star. \cite{Law2009} estimate a 30\% systematic uncertainty in the fluxing, largely from rapid variations of the AO-corrected PSF.
\footnotetext[1]{http://irlab.astro.ucla.edu/osiris/}

\subsection{Magellan/FIRE observations}
\label{subsec:data-fire}
\begin{figure*}
\centering
\includegraphics[width=14cm]{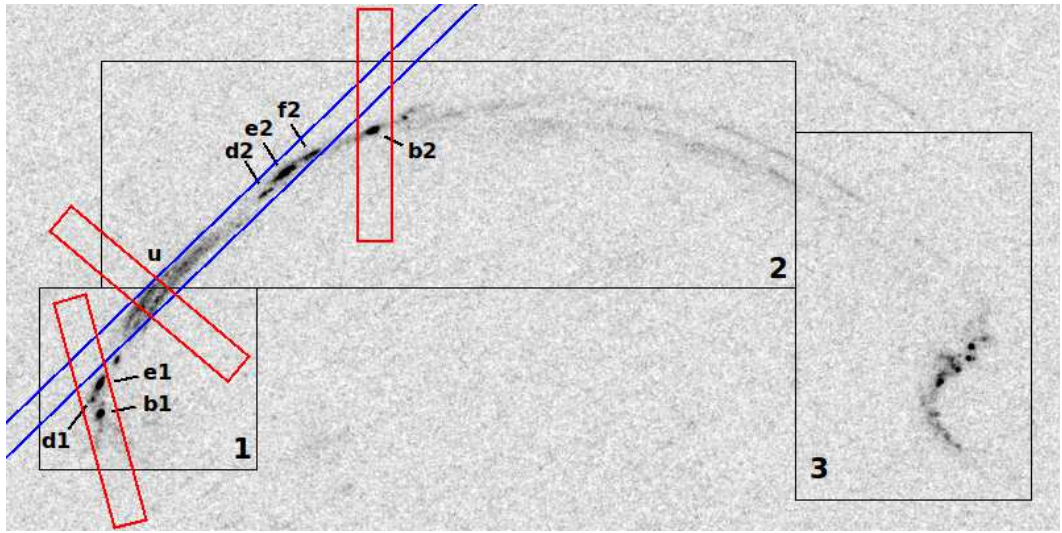}
\caption{32\arcsec\ $\times$ 16 \arcsec\ HST/WFC3 image of RCSGA0327 in the F390W band. The black boxes approximately enclose each of the three images of the source galaxy which together make up the giant arc (as in Figure~\ref{fig:implane}). The $0\farcs76 \times 42$\arcsec\ NIRSPEC slit is shown in blue at PA=134\arcdeg\ East of North. Three pointings of the 1\arcsec\ $\times$ 6\arcsec\ FIRE slit are shown in red, targeting clumps $d1$-$e1$-$b1$ in image 1, clump $u$ and clump $b2$ in image 2 respectively. For simplicity the slits are shown centred on the arc, in reality we placed the source on the left and right sides of the slit for an ABBA nod pattern. \label{fig:slits}}
\end{figure*}

We observed RCSGA0327 with the Folded-Port Infrared Echelette (FIRE; \citealt{Simcoe2013}) at the Magellan Baade telescope in Chile on 2010, October 14-15 UT. The echelle mode delivers a continuous spectrum from 0.82-2.5$\micron$ at a spectral resolution $R=3600$ for the widest 1\arcsec\ $\times$ 6\arcsec\ slit. The seeing was monitored at the telescope and ranged from $0\farcs8$ to $1\farcs2$ on both nights. Based only on ground-based imaging, we chose three separate positions of the arc as shown in Figure~\ref{fig:slits}. With the HST imaging, we now know they correspond to clumps $b1$-$d1$-$e1$ in image 1, clump $u$, and clump $b2$ in image 2. The pointings were acquired through a blind offset from a nearby cluster galaxy; source acquisition was verified with the near-IR slit-viewing camera. The observations consisted of four individual 600~s exposures for each pointing, nodded along the slit in an ABBA pattern. The telluric star HD21875 was observed every hour for flux calibration purposes. We reduced the data using the custom pipeline provided by R. Simcoe \citep{Simcoe2013}, which uses OH skylines for wavelength calibration and performs sky subtraction using the techniques presented by \cite{Kelson2003}. The extracted spectra for clumps $u$ and $b2$ are shown in Figure~\ref{fig:spec}. We cannot spatially resolve clumps $b1$, $d1$ and $e1$ covered by the pointing in image 1. Since clumps $b1$ and $e1$ are of comparable brightness, we cannot derive line fluxes for individual clumps from this pointing and do not consider it further.

For each spectrum, we simultaneously fit all emission lines with a multi-component Gaussian model, using the IDL Levenberg-Marquardt least-squares fitting code MPFITFUN \citep{mpfitfun}. We obtain an initial fit of the bright H$\alpha$ emission line to establish a first guess for the redshift and linewidth. For the combined fit, we set the initial wavelength centroids of all lines based on their NIST rest wavelengths\footnotemark[2] and allow them to vary by up to three times the 1$\sigma$ uncertainty from the initial fit to allow for errors in the wavelength calibration. For each spectrum, all lines are forced to share a common velocity width, since the nebular emission is expected to originate from the same physical region within the galaxy. We report the flux measurements in Table~\ref{tab:fluxes}. In the FIRE spectrum of clump $u$, the [Ne~III]~$\lambda$3869 emission line is not detected. We derive an upper limit as the flux contained within a Gaussian with the common linewidth and a peak value equal to twice the noise at the expected line center. 
\footnotetext[2]{http://www.pa.uky.edu/$\sim$peter/atomic/}

\subsection{Keck/NIRSPEC observations}
\label{subsec:data-nirspec}
\begin{figure}
\centering
\includegraphics[width=8.5cm]{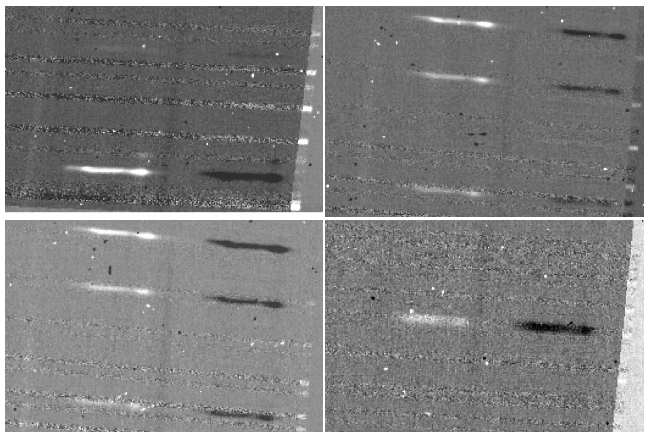}
\caption{Two-dimensional sky-subtracted, nod-subtracted NIRSPEC spectra of RCSGA0327. Each frame is a subtraction of two nods, with one nod as light and the other as dark.  In each panel, wavelength increases from bottom to top.  Counter-clockwise from top left, the panels show: \textit{(top left)}  the N6 spectrum, with H$\alpha$, the [N~II] doublet, and [S~II]; \textit{(top right)} the N3 spectrum, with H$\beta$ and the [O~III]~$\lambda$4959,5007 doublet; \textit{(bottom left)} same as top right, but two additional nods; \textit{(bottom right)} the N1 spectrum, with [O~II]~$\lambda$3727. \label{fig:nirspec2d}}
\end{figure}

\begin{figure*}
\centering
\includegraphics[width=\textwidth,angle=90]{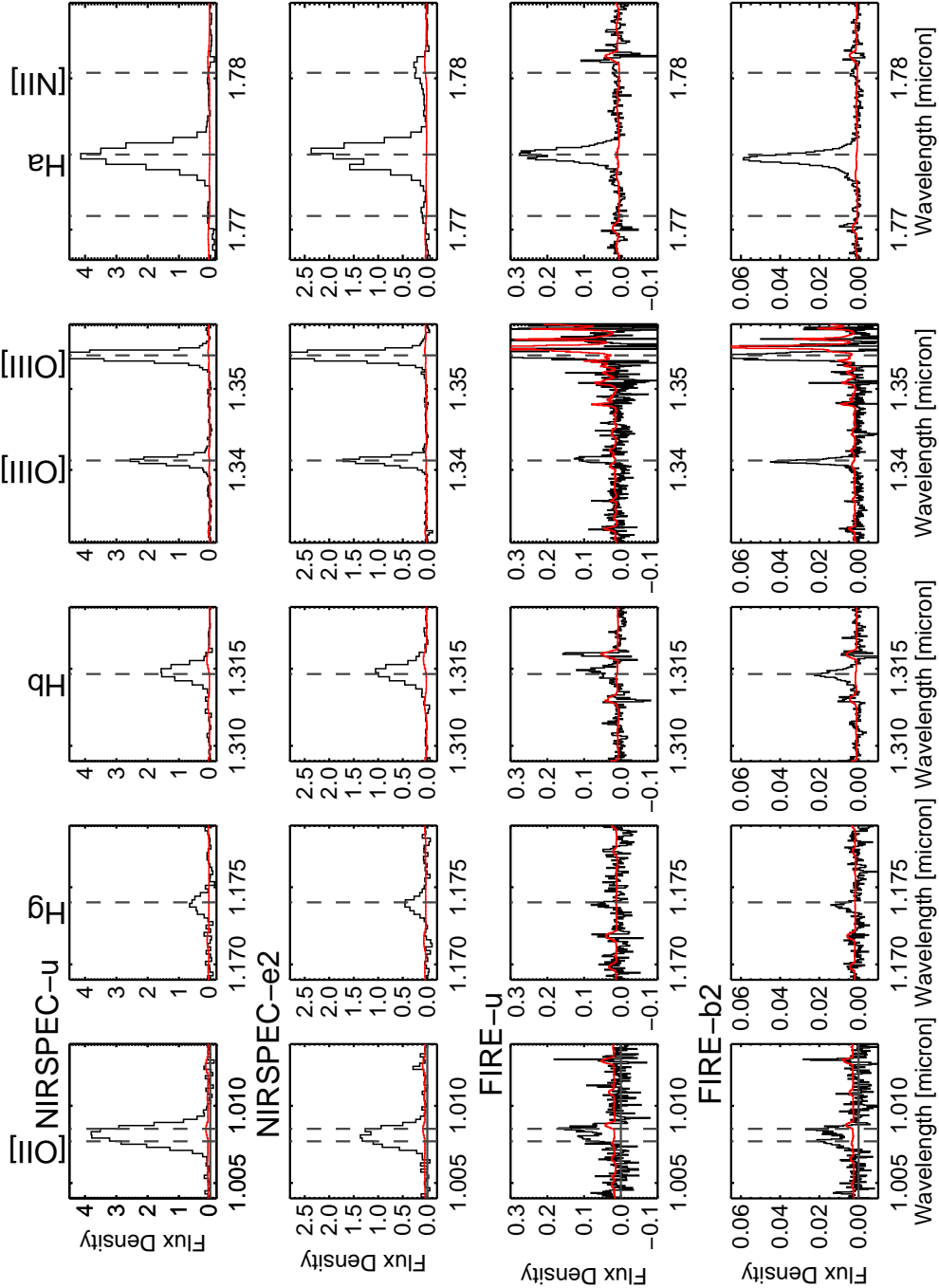}
\vspace{0.15in}
\caption{Extracted NIRSPEC spectra for clumps $u$ and $e2$ and extracted FIRE spectra for clumps $u$ and $b2$. Three wavelength ranges are chosen to show the [O~II]~$\lambda$3727, H$\gamma$, H$\beta$, [O~III]~$\lambda$4959,5007, H$\alpha$ and [N~II] emission lines. The 1$\sigma$ error spectra are shown in red. The y-axis shows specific flux density in units of $10^{-16}$~erg~s$^{-1}$~cm$^{-2}$~\AA$^{-1}$. \label{fig:spec}}
\end{figure*}

We obtained a total of 1.3~hr of near-IR long-slit spectroscopy of RCSGA0327 with Keck/NIRSPEC on 2010 February 4 UT, targeting the brightest 10\arcsec\ of the arc at a PA of 134\arcdeg\ East of North (Figure~\ref{fig:slits}) with three grating settings (NIRSPEC filters N1, N3 and N6). Detailed analysis of the collapsed one-dimensional spectra was published in \cite{Rigby2011}, hereafter R11. The subsequently obtained HST/WFC3 imaging revealed that the slit position covered several individual star-forming regions: clumps $f2$, $e2$ and $d2$ in image 2 and clump $u$ shared between image 1 and 2. Here we reanalyse the spectra to quantify the spatial variation of line fluxes and line ratios across the clumps. 

Details of the observations and the data reduction can be found in R11. Figure~\ref{fig:nirspec2d} shows the two-dimensional spectra for the main emission lines. The spectra are dominated by emission from clumps $u$ and $e2$, which are clearly resolved. Clump $e2$ has a $\sim20$~\% contribution from clumps $f2$ and $d2$, which cannot be separated. To compensate for differential seeing and slit losses, we apply the same bulk scaling of the N1 and N6 filters relative to the N3 filter as in R11. We again note that the relative fluxing should be excellent within a single grating setting, but from one grating setting to another the relative flux offsets may be large. The extracted one-dimensional spectra of clumps $u$ and $e2$ are plotted in Figure~\ref{fig:spec}.  

For each clump and grating setting, we simultaneously fit all emission lines as described above for the FIRE spectra; line fluxes can be found in Table~\ref{tab:fluxes}. Adding line fluxes for clumps $u$ and $e2$ generally agrees within 2$\sigma$ of the line fluxes published in R11. The absolute flux values of clump $u$ should not be compared between the FIRE and NIRSPEC spectra, since we have not applied any aperture correction between both slits. We can address the relative fluxing of the different gratings by comparing line ratios for clump $u$ to the ratios measured in the FIRE spectrum of this clump. Using [O~II]~$\lambda$3727 in N1, H$\gamma$, H$\beta$ and [O~III]~$\lambda$4959 in N3, and H$\alpha$ in N6, we find a weighted mean offset of 9\%, 35\% and 40\% for N6-N3, N3-N1 and N6-N1 respectfully. We caution that the only bright line in N1, the [O~II]~$\lambda$3727 doublet, is difficult to measure accurately.

\begin{deluxetable}{lll|ll}
\tabletypesize{\footnotesize}
\tablewidth{0pc}
\tablecaption{Measured line fluxes from NIRSPEC and FIRE. \label{tab:fluxes}}
\tablehead{
\colhead{} &  \multicolumn{2}{c}{NIRSPEC} & \multicolumn{2}{c}{FIRE} \\
\colhead{emission line} &  \colhead{clump $u$} & \colhead{clump $e2$} &  \colhead{clump $u$} & \colhead{clump $b2$}}
\startdata
$[$O~II]~$\lambda$3727    & $544\pm13$ & $189\pm8$ & $12.6\pm1.1$ & $2.3\pm0.2$     \\
$[$Ne~III]~$\lambda$3869 &   $64\pm5$   &   $30\pm3$ & $<1.8$          & $0.69\pm0.05$ \\
H$\gamma$      &   $68\pm3$   &  $50\pm2$  & $2.4\pm0.3$   & $0.64\pm0.04$ \\   
H$\beta$           & $171\pm3$   & $119\pm2$  & $5.8\pm0.3$  & $1.40\pm0.04$  \\
$[$O~III]~$\lambda$4959       & $275\pm2$  & $189\pm2$  & $10.1\pm0.5$  & $2.99\pm0.07$  \\
$[$O~III]~$\lambda$5007       & $909\pm4$  & $611\pm3$  & $44\pm1$       & $7.28\pm0.12$  \\
H$\alpha$         & $675\pm3$   & $388\pm2$ & $26.1\pm0.5$ & $4.82\pm0.06$ \\
$[$N~II]~$\lambda$6583        &   $34\pm4$  &   $51\pm3$ & $1.3\pm0.2$  & $0.20\pm0.02$ \\
\enddata
\tablecomments{Fluxes are in units of  $10^{-17}$~erg~s$^{-1}$~cm$^{-2}$.}
\end{deluxetable}

\section{Galaxy Kinematics}
\label{sec:kin}
\subsection{Mapping the Velocity Field}
\label{subsec:kinmaps}[h]

\begin{figure*}
\centering
\vspace*{-10cm}
\includegraphics[width=17cm]{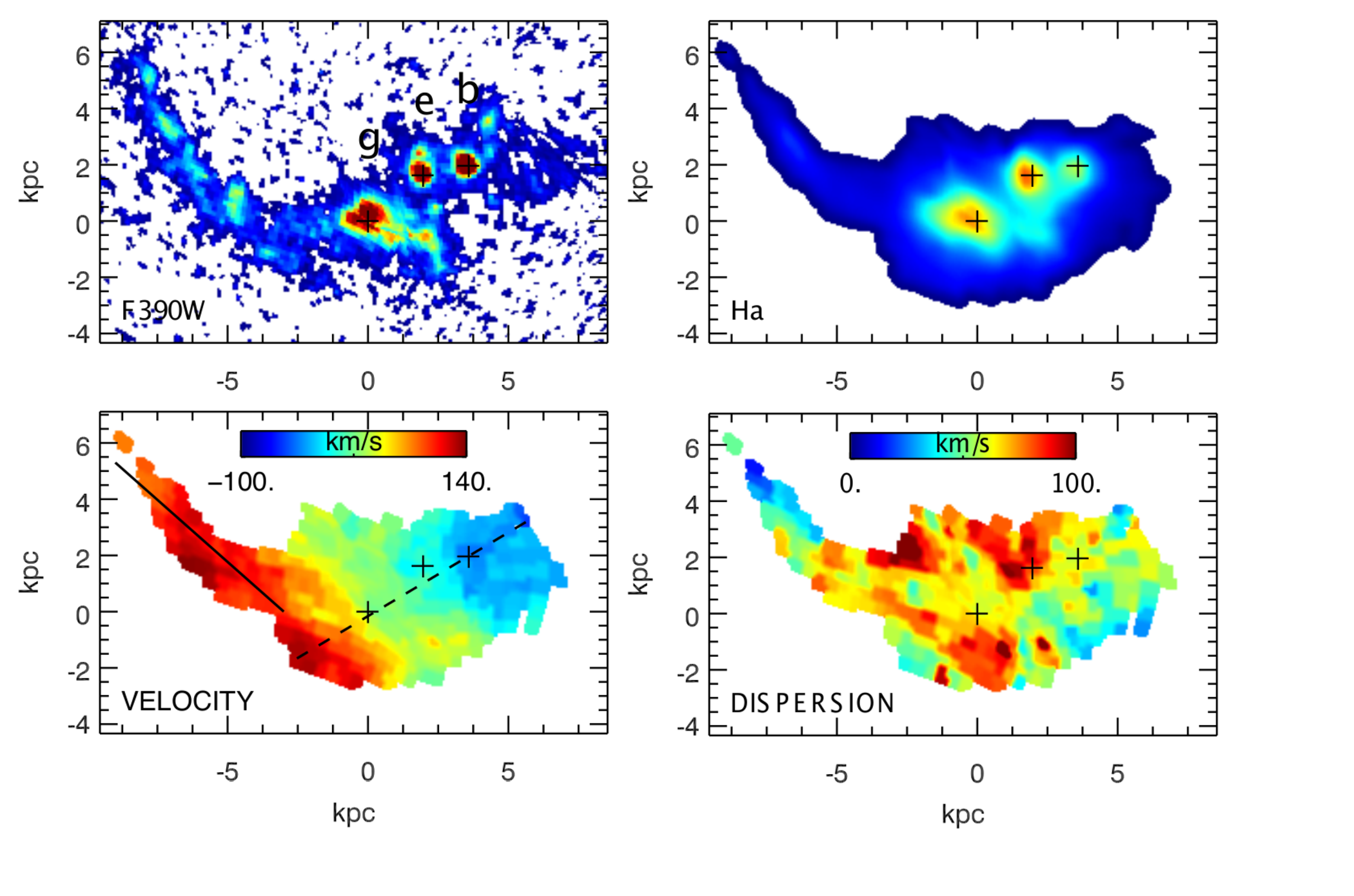}
\vspace{-0.8cm}
\caption{Source-plane maps of the F390W flux, H$\alpha$ flux, velocity, and velocity dispersion for image 3. The latter three maps are smoothed with a boxcar average of 3 pixels for the purpose of visualisation.The x- and y-axes are centered on clump $g$; clumps $g$, $e$, and $b$ are marked by black crosses (from left to right). The kinematic axis and the axis along the  ``arm'' which extends to the North-East from clump $g$ are overlaid on the velocity map as dashed and solid lines respectively. \label{fig:kin3}}
\vspace{-10cm}
\includegraphics[width=17cm]{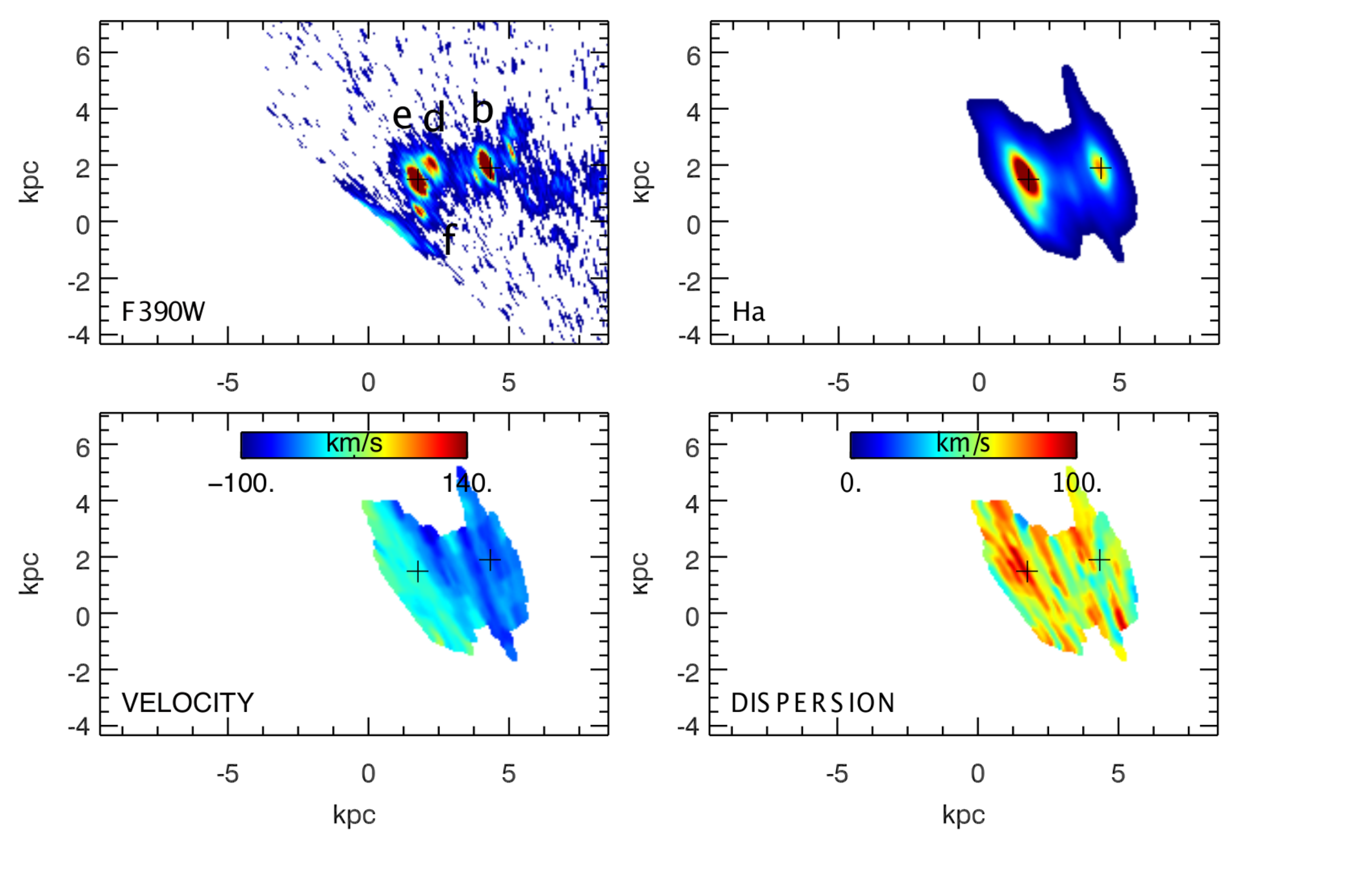}
\vspace{-0.8cm}
\caption{Source-plane maps for image 2, which contains a smaller, but more highly-magnified part of the source-plane galaxy. The size, centering and scaling of the maps is identical to Figure~\ref{fig:kin3}. Clumps $e$ and $b$ are marked by black crosses (from left to right). \label{fig:kin1}}
\end{figure*}
We create spatial and kinematic maps of the H$\alpha$ emission in RCSGA0327 by fitting a Gaussian profile to the H$\alpha$ emission line for every spatial pixel. The noise is estimated separately from a blank region of sky in each of the pointings. A minimum signal-to-noise of $4.5\sigma$ is required for a detection of H$\alpha$; if this criterion is not met, the surrounding $3 \times 3$ spatial pixels are averaged and the fit is re-attempted. Formal uncertainties are derived for each spatial pixel from fitting 100 mock spectra consistent with the noise.
The relative shifts of the wavelength centroid of the H$\alpha$ emission line translate into a velocity map of the ionized gas. We define an absolute velocity zeropoint based on the H$\alpha$ centroid determined for clump $e$ from the NIRSPEC spectra (\S~\ref{subsec:data-nirspec}), which corresponds to $\lambda_{H\alpha} = 1.774903\pm0.000003$\micron\ or $z=1.7037455\pm0.000005$. After applying a barycentric correction to the OSIRIS wavelength calibration, we find a wavelength shift of 1.1\AA\ for pointing 2 and 0.7\AA\ for pointing 3, well within the OSIRIS wavelength calibration uncertainty of up to 5\AA \footnotemark[3]. We correct the calibration of both pointings for these wavelength shifts. 
\footnotetext[3]{http://irlab.astro.ucla.edu/osiriswiki/}

The H$\alpha$ flux, velocity and velocity dispersion maps are transformed into the source plane and shown in Figures~\ref{fig:kin3} and \ref{fig:kin1}. The maps are smoothed with a boxcar average of 3 pixels for the purpose of visualisation. The velocity dispersion is corrected for the instrument response function by subtracting the instrumental resolution in quadrature. This is determined from Gaussian fits to the OH sky lines and corresponds to a FWHM of 5.6\AA. Typical uncertainties in the velocity and velocity dispersion maps are $\sim10$~km/s. The x- and y-axes of the maps have been centered on the brightest clump, clump $g$. 
To allow the source-plane transformations, the OSIRIS maps have to be aligned with the HST images on which the lens model is based. This is done visually, using the positions of the various clumps in both datasets. While this alignment does not produce global spatial accuracy to better than one OSIRIS pixel ($0\farcs1$), it does not influence the comparison of the relative location of multiple emission features within each pointing. 

The velocity map agrees with the velocity offsets between clumps measured in the FIRE and NIRSPEC data. Based on the H$\alpha$ emission line centroid, we find $\delta v = 15\pm1$~km~s$^{-1}$ between clumps $u$ and $e2$ from the NIRSPEC spectra and $\delta v = 44\pm3$~km~s$^{-1}$ between clumps $u$ and $b2$ from the FIRE spectra. 

\subsection{Analyzing the Velocity Field} 
\label{subsec:kinan}
\begin{figure}
\centering
\includegraphics[width=9cm]{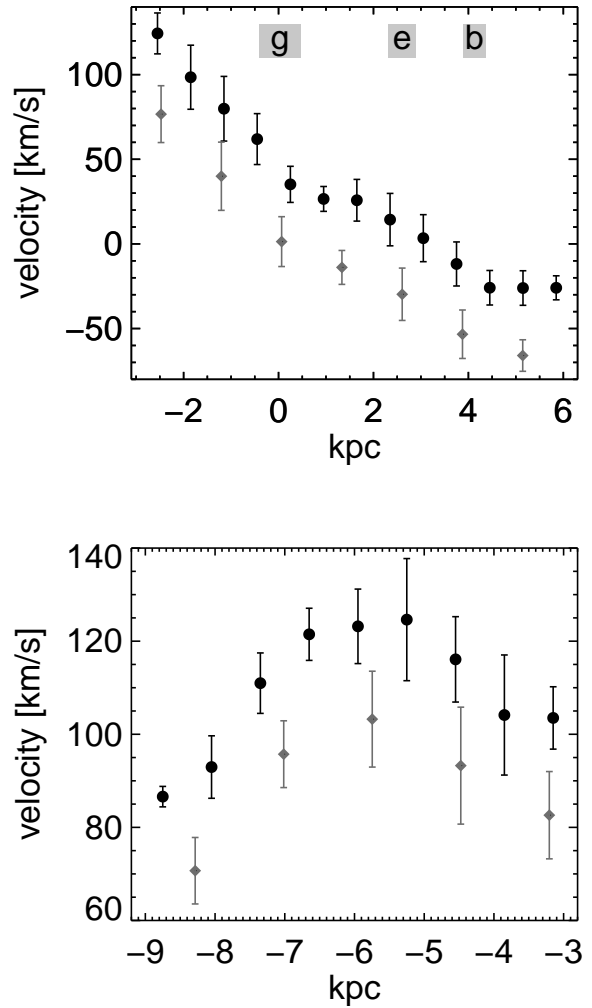}
\caption{\textit{(Top)} 1D velocity profile of the source-plane galaxy along its kinematic axis, constructed from the median and standard deviation of the velocity map in 0.7~kpc wide bins. 
The locations of clumps $g$, $e$ and $b$ are indicated for reference. \textit{(Bottom)} 1D velocity profile along the ``arm'' extending towards the North-East from clump $g$, indicated by the solid line in Figure~\ref{fig:kin3}. The peak in velocity near the middle of the arm is typical for a tidal tail. In both panels, the gray diamonds indicate the coarser spatial resolution available without lensing magnification; the points are shifted down in velocity by 40~km/s and 20~km/s in the top and bottom panels respectively to improve the clarity of the figure.\label{fig:kin1d}} 
\end{figure}

The velocity map shown in Figure~\ref{fig:kin3} is well-structured and contains a strong velocity gradient. We define a kinematic axis connecting the maximum and minimum velocity regions and extract a 1D velocity profile as the median and standard deviation within bins of 0.7~kpc width along this axis (Figure~\ref{fig:kin1d}; top panel). The bins were chosen in correspondence with the FWHM of the OSIRIS datacubes such that individual datapoints in the 1D profile represent independent measurements. The locations of clumps $g$, $e$ and $b$ are shown for reference. This 1D profile does not display the smooth S-shaped curve expected from a rotating disk but instead levels off to a plateau between clumps $g$ and $e$. This suggests the system is currently undergoing an interaction. The bottom panel in Figure~\ref{fig:kin1d} shows the 1D velocity profile extracted along the solid line in Figure~\ref{fig:kin3} (PA=40\arcdeg\ East of North). This corresponds to a region of clumpy emission (clumps $h$ through $k$, see Figure~\ref{fig:implane}) which extends from clump $g$ towards the North-East and shows blue broad-band rest-frame UV colors. The velocity profile agrees with the expectation for a tidal tail swinging away from the observer and curving back into the plane of the sky: the velocity peaks in the middle and falls off to either side. Within this interpretation, separate star-forming knots in the tail (clumps $h$-$k$) might evolve into tidal dwarf galaxies (e.g. \citealt{Mihos1998, Hibbard1994, Duc2000}). 

The gray diamonds in Figure~\ref{fig:kin1d} (shifted down in velocity to improve the clarity of the figure) illustrate the coarser spatial resolution that would be obtained without lensing magnification. The velocity peak in the tidal tail remains visible, but the plateau between clumps $g$ and $e$ would be largely smoothed out, resulting in an ambiguous velocity profile which could easily be interpreted as a single rotating disk. 
\npar
The velocity dispersion maps contain additional kinematic information. For a rotating disk at higher redshift, the dispersion is typically seen to increase towards the center due to beam-smearing of the velocity gradient (e.g. \citealt{Epinat2010}). Additionally, areas of active star formation, like clumps, can show elevated velocity dispersion because of outflows \citep{Newman2012a}. RCSGA0327 shows a noticeable peak in velocity dispersion at the location of clump $e$, which agrees with the detection of a strong outflow from this clump presented in \S~\ref{subsec:clumposiris}. The other two peaks in velocity dispersion are located between clump $g$ and the tidal tail and likely originate from increased turbulence due to the ongoing interaction. Additionally, elevated velocity dispersions could be caused by overlapping components along the line of sight, where the H$\alpha$ profile consists of two emission lines separated slightly in velocity. Attempting to fit the H$\alpha$ emission with a single Gaussian function will result in an overestimate of the linewidth. This has been seen in local mergers (e.g. \citealt{Mihos1998}), but we lack sufficient signal-to-noise to robustly identify any such multiple H$\alpha$ emission peaks in the OSIRIS data.
\npar
The physical extent of RCSGA0327 further strengthens the kinematic arguments for an ongoing interaction. We estimate the size of the system from a segmentation map of all pixels $>3\sigma$ in the source-plane F160W image, including the arm which extends to the North-East. With a radius $r_e=\sqrt{A/\pi}=7.1$~kpc and a stellar mass of $6 \times 10^9$~M$_\odot$ \citep{me2012a}, RCSGA0327 lies near the upper extreme of the size-mass relation of star-forming galaxies at $z\sim2$, even when taking into account the considerable scatter in this relation \citep{Franx2008,Williams2010,Wuyts2011,Barro2012}. Thus, as an isolated galaxy this system would be unusually large for its stellar mass.

\section{Physical Properties of Clumps}
\label{sec:clumps}
This section covers measurements of the individual star-forming regions that can be identified in RCSGA0327. From the WFC3 imaging, we measure broad-band photometry and use spectral energy distribution (SED) modelling to constrain the stellar populations of the clumps: age, stellar mass, extinction and SFR. Integrated H$\alpha$ spectra are created for each clump from the OSIRIS data, from which velocity dispersion, SFR, metallicity and outflow properties can be derived.

\subsection{Broad-band Photometry}
\label{subsec:clumpphot}
The photometric analysis of the clumps is performed in the image plane because the PSF is not well defined in the source-plane reconstructions. It varies across the source plane depending on the location with respect to the lensing caustic, and has an elliptical shape due to the non-isotropic magnification (see \citealt{Sharon2012} for more details). The clumps are identified in the higher resolution WFC3/UVIS images; we transform the WFC3/IR images to the same reference for a uniform photometry measurement.
A series of elliptical apertures of increasing radial extent is defined for each region. The ellipticity changes according to how significantly the clumps are stretched by the lensing. We measure the flux at a radius of roughly twice the FWHM and use the zeropoint and aperture corrections (typically 15-25\%) as defined in the WFC3 handbook\footnotemark[4]. 
\footnotetext[4]{http://www.stsci.edu/hst/wfc3/phot\_zp\_lbn} 

The clumps are embedded in the galaxy, and the removal of underlying galaxy background light has to be handled carefully. At rest-frame UV wavelengths, the contribution of the diffuse galaxy background is often ignored since the clumps are typically 2-4 times as bright as their surroundings \citep{Elmegreen2009}. However, the lensing magnification of RCSGA0327 allows us to study smaller and fainter clumps; we find that the galaxy background accounts for 30-80\% of the total flux in the clump apertures. Additionally, the contrast between the clumps and the background is lower at rest-frame optical wavelengths, where the emission of bright, young stars is less dominant. Following \cite{Guo2012}, we determine the background for each WFC3 band as all pixels that belong to the galaxy (i.e. lie above a threshold of 3$\sigma$) and do not fall within one of the photometric clump apertures. The \textit{global} background is then simply the average value of these background pixels. We also define a \textit{local} background for each clump as the mean and standard deviation of the background pixels within an annulus of width $\sim0.4$\arcsec\ outside the clump aperture. For RCSGA0327, the local backgrounds generally agree with the global estimate. Similarly, \cite{Guo2012} find that the change in rest-frame UV colors between a global and local background subtraction for $z\sim2$ SFGs in the Hubble Ultra Deep Field is not statistically significant. However, it remains important to check this consistently in future studies since an incorrect assumption of a constant global background across the galaxy can introduce false radial variations in clump color and related properties. 

\begin{figure}
\centering
\includegraphics[width=9cm]{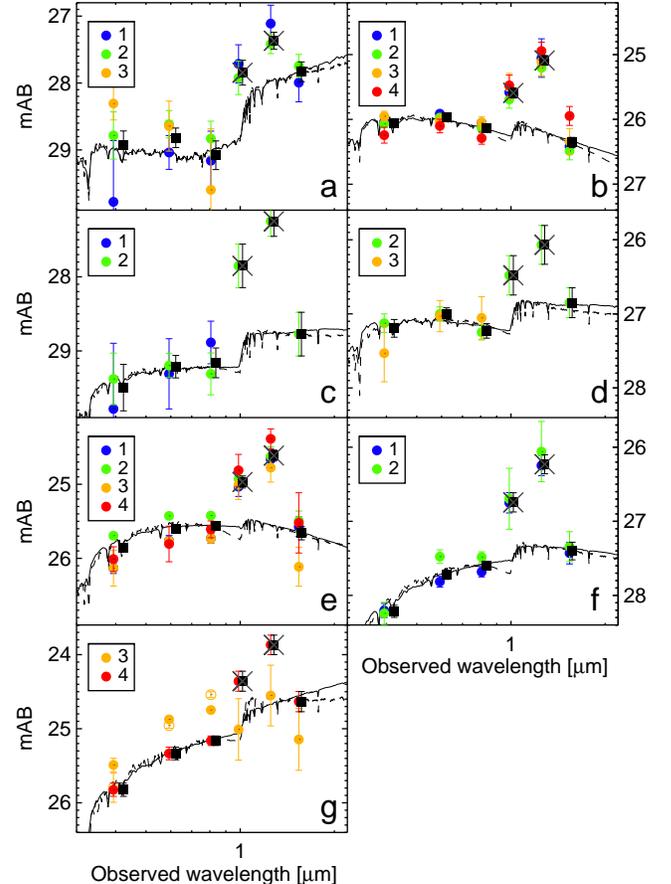}
\caption{Final photometry for the clumps as measured from the six WFC3 UVIS and IR bands, corrected for the lensing magnification. The measurements from the multiple images of the source are in reasonable agreement. The black squares correspond to the weighted mean and uncertainty in the mean of the photometry for each clump; they are shifted slightly in wavelength to improve the clarity of the figure. The F098M and F125W magnitudes are not corrected for nebular line emission, they are shown with large grey crosses in all panels and not included in the SED fit. The best-fit SED to the weighted mean UVIS+F160W photometry is shown in black, with a solid line for the default Calzetti extinction law and a dashed line for the SMC extinction law. \label{fig:clumpmag}}
\end{figure}

Outliers within the clump apertures, such as contributions from neighbouring clumps, are masked by hand. Identifying exactly which pixels should be masked can be uncertain in the IR images where the larger PSF blends close overlapping neighbours such as clumps $d$, $e$ and $f$ in image 2 and clumps $d$, $e$ and $b$ in image 3. In those cases, the added flux uncertainty is typically 10\%. In image 3, the two cluster galaxies that fall on top of the arc are removed before measuring the clump photometry, using the technique presented in \cite{me2010}. In short, for each WFC3 band we subtract a scaled F390W image to remove the arc. Any remaining positive flux at the positions of the cluster galaxies is then subtracted and the original frame is restored by adding the scaled F390W image back in. 
Final magnitudes are corrected for galactic extinction \citep{Schlegel1998}. The photometric uncertainties include Poisson noise, an absolute WFC3 zeropoint uncertainty of 1\%, an uncertainty in the background subtraction determined from the standard deviation of the local background value, and an uncertainty from the masking of neighbouring clumps. The background subtraction dominates the total uncertainty.
\npar
For the galaxy-integrated photometry measurement of RCSGA0327, the contribution of rest-frame optical nebular emission lines to the near-IR photometry was estimated at 5-10\%, which is negligible compared to the photometric uncertainties \citep{me2012a}. However, individual star-forming regions have much higher star formation surface densities than the host galaxy as a whole, such that the nebular emission adds a sizeable, and possibly dominant, contribution to the near-IR light. At the redshift of RCSGA0327, the F098M band includes the [O~II]~$\lambda$3727 emission line, while H$\beta$ and the [O~III]~$\lambda$4959,5007 doublet are the most important contaminating lines for the F125W band. H$\alpha$ falls just redward of the F160W band. The removal of this nebular line emission from the broad-band photometry is not trivial. We can use the F132N narrowband image to measure the H$\beta$ line emission within each of the clump apertures used for the broad-band photometry. Knowledge of the line ratios of H$\beta$ to the [O~II] and [O~III] emission lines then allows removal of the line contamination from the F098M and F125W fluxes. However, the NIRSPEC and FIRE data show that these line ratios can vary across the arc by more than a factor of two (Table~\ref{tab:fluxes}). Correcting individual clumps for emission line contamination therefore adds large uncertainties to the F098M and F125W photometric uncertainty, which results in minimal constraints on the observed spectral energy distribution. For this work, the F098M and F125W bands remain uncorrected and are not included in the SED modelling described below. 
\npar
Finally, the photometry for each clump needs to be corrected for the lensing magnification. Using the magnification map presented in \cite{Sharon2012}, we derive individual clump magnifications as the flux-weighted mean magnification within the photometric clump aperture. Figure~\ref{fig:clumpmag} shows the demagnified AB magnitudes for each of the clumps in the multiple images of the source galaxy. Not all clumps are present in all images and some of the clumps are too faint and/or blended for an accurate measurement, especially in the WFC3/IR bands. There is a general agreement between the independent clump measurements from the multiple images, which confirms the robustness of the photometry method. The weighted mean and uncertainty in the mean derived from the available measurements for each clump are shown with black datapoints. We use this weighted mean in all subsequent analysis. Image 3 is complicated by the presence of cluster member $G1$ (see Figure~\ref{fig:implane}), which appears to split clump $g$ in two images. The photometry of those two images (shown with open and filled orange circles in the bottom left panel of Figure ~\ref{fig:clumpmag}) does not agree with the well-measured appearance of clump $g$ in image 4. Given this lensing complexity, we discard the measurements of clump $g$ from image 3.

\subsection{SED Modelling}
\label{subsec:clumpsed}
The stellar populations of the clumps can be constrained with spectral energy distribution modelling of the observed clump photometry. We use the SED fitting code FAST \citep{Kriek2009} at fixed spectroscopic redshift with the \cite{Bruzual2003} stellar population synthesis models (BC03), a \cite{Chabrier2003} IMF and \cite{Calzetti2000} dust extinction law. The metallicity is restricted to 0.2 or 0.4\,Z$_\odot$ for all clumps, consistent with the oxygen abundance measured from the integrated NIRSPEC spectrum (R11), as well as the metallicity measurements of individual clumps from the OSIRIS, FIRE and NIRSPEC data (see \S~\ref{subsec:clumposiris}). We initially adopt exponentially decreasing star formation histories (SFH) with minimum $e$-folding time $\log(\tau)=8.5$ and a minimum age limit of 50~Myr. \cite{Wuyts2011} shows good agreement between SFR estimates based on these assumptions and other multi-wavelength SFR indicators out to $z\sim3$. This default fit returns best-fit models at the age cut-off of 50~Myr for all clumps except clump $a$. Removing the age limit significantly improves the fit and returns ages between 3 and 16~Myr, albeit with large uncertainties. The 50~Myr age limit roughly corresponds to the dynamical time scale of a $z\sim2$ galaxy and is typically included to avoid a luminosity bias from the most recently formed O and B stars in the SED fit. However, star-forming clumps have much shorter dynamical time scales ($t_{dyn} \sim r/v \sim 5$~Myr) and in the local Universe, star clusters are often dated between a few and a few ten Myr (e.g. \citealt{Bastian2006}). We caution that these age estimates should be interpreted as the age of the current episode of star formation within the regions, which outshines the contribution of a possible underlying older stellar population (age $>100$~Myr).

The best-fit SED models for the weighted mean photometry of each clump are shown in Figure~\ref{fig:clumpmag} and 68\% confidence intervals for the stellar population parameters are reported in Table~\ref{tab:clumpsed}. The stellar masses fall below the typical clump masses of $10^8-10^{10}$~M$_\odot$ reported in non-lensed studies \citep{Forster2011, Guo2012}, and lie closer to the range of $10^6-10^{8}$~M$_\odot$ found for a lensed spiral galaxy at $z=1.5$ \citep{Adamo2013}.
As is often seen in SED modelling, the $e$-folding time $\tau$ is not constrained by the fit. 
\npar
The SFRs derived from the best-fit models are very high, especially for clumps $e$ and $g$ with 110 and 60~M$_\odot$~yr$^{-1}$ respectively. From a range of different SFR indicators, \citet{me2012a} estimate a galaxy-integrated SFR for RCSGA0327 of 30-50~M$_\odot$~yr$^{-1}$. This suggests that the SED fit is overestimating the SFR of individual clumps. We experimented with alternative SFHs by including a) exponentially declining models with shorter $e$-folding times $\tau$ down to 10~Myr; b) delayed histories with $SFR \sim t \exp(-t/\tau)$; and c) inverted models with $SFR \sim \exp(t/\tau)$. The stellar population parameters remain within the $1\sigma$ uncertainties of the default fit for all these alternative SFH templates. At the young clump ages of $\lesssim20$~Myr, the shape of the SFH has negligible impact on the best-fit stellar population parameters and is therefore not causing the high clump SFRs.

The SED-derived SFR estimates can be reduced by assuming a different dust extinction law. It has been suggested that the assumption of a patchy dust distribution inherent in the Calzetti dust extinction law might not be a good representation of the dust geometry in young star-forming galaxies at $z\sim2$ and could significantly overpredict their dust extinction \citep{Reddy2006,Siana2009,me2012a}. The large covering fraction of outflowing gas observed for two $z\sim2$ lensed SFGs inferred from the presence of opaque interstellar absorption lines in their rest-frame UV spectra \citep{Siana2008,Siana2009}, is indicative of a more uniform foreground sheet of dust. This results in a steeper extinction law, such as the one derived for the Small Magellanic Cloud \citep{Prevot1984}. Rest-frame UV spectra of clumps $e2$ and $u$ in RCSGA0327 taken with the MAGE spectrograph at Magellan, show similar opaque absorption lines (J.~R.~Rigby et al. 2014, in preparation). When adopting the SMC extinction law in the SED fit, we cannot distinguish the best-fit models from the Calzetti result in terms of $\chi^2$-statistics. The best-fit models and derived stellar population parameters are included in Figure~\ref{fig:clumpmag} and Table~\ref{tab:clumpsed}. The SFRs of clumps $e$ and $g$ are now 5 and 3.1~M$_\odot$~yr$^{-1}$ respectively, a more plausible result in light of both the galaxy-integrated SFR and the H$\alpha$-derived clump SFRs (see \S~\ref{subsec:clumposiris}). The reddening is lower due to the steeper extinction curve, and the stellar ages are overall higher than for the Calzetti fit. The stellar mass is consistent within the $1\sigma$ uncertainties and therefore not included in Table~\ref{tab:clumpsed}.

\begin{deluxetable*}{lcccc|ccc|c}
\tabletypesize{\footnotesize}
\tablewidth{0pc}
\tablecaption{Clump Stellar Population Parameters and Radii. \label{tab:clumpsed}}
\tablehead{
\colhead{} & \multicolumn{4}{c}{Calzetti extinction} & \multicolumn{3}{c}{SMC extinction} & \colhead{} \\                            
\colhead{} & \colhead{Age} & \colhead{$E(B-V)_s$} & \colhead{$\log(M_*/\mathrm{M}_\odot)$} & \colhead{$SFR$} & \colhead{Age} & \colhead{$E(B-V)_s$} & \colhead{$SFR$} & \colhead{$r_{cl}$}  \\
\colhead{} & \colhead{(Myr)} & \colhead{} & \colhead{} & \colhead{(M$_\odot$~yr$^{-1}$)} & \colhead{(Myr)} & \colhead{} & \colhead{(M$_\odot$~yr$^{-1}$)} & \colhead{(pc)}} 
\startdata
$a$ & $3200^{+0}_{-2800}$ &  $0.02^{+0.13}_{-0.02}$ & $8.1^{+0.1}_{-0.5}$ & $0.05^{+0.10}_{-0.02}$ & $2500^{+1250}_{-1350}$ & $0.00^{+0.04}_{-0.00}$ & $0.05^{+0.02}_{-0.01}$ & 250$\pm$120 \\  
$b$ &       $6^{+21}_{-5}$    &  $0.17^{+0.10}_{-0.10}$ & $7.7^{+0.5}_{-0.2}$ &    $8^{+124}_{-6}$ &     $8^{+1}_{-2}$ &    $0.07^{+0.01}_{-0.01}$ & $3.3^{+0.9}_{-1.0}$   & 180$\pm$100 \\
$c$ &     $10^{+1140}_{-9}$ &  $0.25^{+0.2}_{-0.25}$   & $6.8^{+0.8}_{-0.2}$ &    $1^{+24}_{-1}$ & $90^{+420}_{-70}$ & $0.07^{+0.04}_{-0.07}$ & $0.05^{+0.09}_{-0.04}$ & 270$\pm$160 \\
$d$ &     $16^{+573}_{-13}$ &  $0.17^{+0.14}_{-0.17}$ & $7.5^{+0.6}_{-0.2}$ &    $2^{+12}_{-2}$ & $100^{+220}_{-90}$ & $0.04^{+0.04}_{-0.00}$ & $0.4^{+0.3}_{-0.2}$   & 320$\pm$190 \\
$e$ &       $3^{+1}_{-2}$      &  $0.32^{+0.03}_{-0.08}$  & $8.5^{+0.1}_{-0.4}$ &       $110^{+220}_{-90}$ &    $11^{+9}_{-5}$ & $0.11^{+0.04}_{-0.00}$ &  $5.0^{+8.3}_{-1.0}$   & 170$\pm$120 \\
$f$ &        $5^{+2}_{4}$       &  $0.37^{+0.08}_{-0.06}$  & $7.7^{+0.4}_{-0.2}$ &    $9^{+91}_{-4}$ &    $22^{+18}_{-14}$ & $0.15^{+0.00}_{-0.04}$ & $0.6^{+0.4}_{-0.4}$   & 160$\pm$80 \\
$g$ &       $8^{+65}_{-7}$    &  $0.37^{+0.13}_{-0.12}$  & $8.7^{+0.5}_{-0.1}$ &       $60^{+1320}_{-50}$ &  $64^{+116}_{-28}$ & $0.11^{+0.04}_{-0.00}$ &  $3.1^{+2.1}_{-2.4}$   & 200$\pm$140 \\
\enddata
\end{deluxetable*}

\subsection{Clump Size}
\label{subsec:clumpsize}
Accurately determining the size of the individual clumps is not trivial. In the literature, region size is often measured from the area above a chosen surface brightness level. This isophote method has a few important problems. First of all, the surface brightness threshold is typically determined visually, from a trade-off between identifying a maximum number of regions while minimizing blending between individual regions. The chosen isophote is thus subjective and difficult to compare between studies, especially over a range of redshifts. Secondly, this method can be influenced significantly by local background variations, especially at high redshifts where undetected low surface brightness regions or overlap light from regions that are only separated by a few pixels can enhance the local background level. Finally, and most importantly, the isophote method will automatically find brighter regions to be larger, since more of the diffuse outskirts of the region will fall above the chosen isophote.

A more robust way to measure region size is provided by the core method, in which a 2D light profile is fitted to the surface brightness profile of each region (e.g. \citealt{Wisnioski2012}). The local background is a free parameter in the profile fit, thus minimizing its influence on the size measurement. Most commonly, a 2D Gaussian light profile is used, which probes primarily the central ionized core of the H~II regions. Adopting this method, we use GALFIT (version 3.0, \citealt{Peng2010}) to create a model for each individual star-forming region in the F390W image, using a Tiny Tim model for the PSF \citep{tinytim}. The F390W image is chosen for the profile fit because the contrast between the clumps and the background is largest in this bluest HST band. A single Gaussian profile provides an adequate fit for most regions, except for the brightest ones, clumps $b$, $e$ and $g$. There, an additional Gaussian component is required to model the surrounding diffuse nebula, which is bright enough to rise above the background level. The best-fit GALFIT models are deconvolved and mapped back to the source plane. 
We define the clump radius $r_{cl}$ as the effective 1$\sigma$ Gaussian width of the source-plane model. 
Table~\ref{tab:clumpsed} reports the weighted mean of the radii measured for the multiple images of each clump. The uncertainties of 50-70~\% reflect the scatter in size measurements for the same clump in the multiple images of the source as well as a systematic uncertainty of 30\%, to account for uncertainties introduced when the clumps are not truly Gaussian as well as resolution effects \citep{Wisnioski2012}. 

The clumps in RCSGA0327 range in size from $\sim$300 to 600~pc (quoted here as the clump diameter). Clumps studied in other lensed galaxies in the literature occupy a similar size range of $\sim300-1000$~pc \citep{Swinbank2009, Jones2010, Livermore2012}. This is significantly smaller than the clump sizes of 1-2~kpc typically measured for non-lensed galaxies (e.g. \citealt{Genzel2011, Wisnioski2012}), where the limited available resolution can blend neighbouring clumps into larger, more luminous regions.
Additionally, since lensing studies typically target less massive galaxies, they are expected to probe less massive, smaller clumps. 

\subsection{Clumps in OSIRIS}
\label{subsec:clumposiris}
\begin{figure}
\centering
\includegraphics[width=8cm]{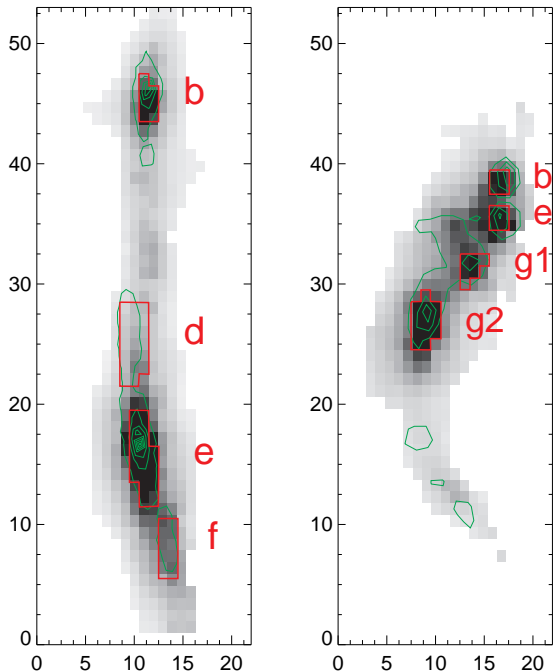}
\caption{Image-plane H$\alpha$ emission maps for pointing 2 \textit{(left)} and pointing 3 \textit{(right)}, overlayed with F390W flux contours in green. The maps show the total H$\alpha$ flux in each spatial pixel in the OSIRIS data cubes as derived from Gaussian fits to the line profile (see \S~\ref{subsec:kinmaps}). The clump apertures are shown in red. These result from the simultaneous fit of multiple 2D Gaussians to the H$\alpha$ emission maps and correspond to the 1$\sigma$ extent of the best-fit 2D Gaussians. \label{fig:clumps}}
\end{figure}

The OSIRIS data provide additional information on the individual star-forming regions. To avoid the ill-defined source-plane PSF as well as uncertainties introduced when mapping the OSIRIS data to the HST reference frame and subsequently to the galaxy source plane, the clumps are identified and characterized in the image-plane datacubes. Implementing the core-method for region identification, we fit multiple 2D Gaussian profiles simultaneously to the H$\alpha$ intensity maps. This identifies clumps $b$ and $e$ in both pointings, as well as both appearances of clump $g$ in pointing 3. Clump apertures are defined as the 1$\sigma$ extent of the best-fit Gaussians. Through a comparison with the F390W image, we additionally identify clumps $d$ and $f$ in pointing 2. These two clumps can also be seen in pointing 3, but there they only extend over one or two pixels, which introduces large uncertainties.
Figure~\ref{fig:clumps} shows the image-plane H$\alpha$ emission maps for both pointings, overlayed with the F390W contours in green. The clump apertures are marked in red. There is good agreement between the rest-frame UV and H$\alpha$ morphology.

An integrated spectrum is constructed for each clump by adding all the spatial pixels within its aperture. We note that typical rotation signatures across individual clumps are insignificant compared to the velocity uncertainties; shifting all pixels to a common central wavelength to correct for velocity broadening has negligible influence on the shape of the integrated spectra.
As discussed above, the core method provides a more robust region identification compared to the subjective isophote method. The total H$\alpha$ emission will be somewhat underestimated due to diffuse emission beyond the clump aperture, which is aggravated by an imperfect Strehl ratio for the adaptive optics correction. 
The wavelength centroid and linewidth do not depend critically on the clump aperture.

\begin{deluxetable*}{lcccccccc}
\tabletypesize{\footnotesize}
\tablewidth{0pc}
\tablecaption{Clump Properties from the OSIRIS Data. \label{tab:clumposiris}}
\tablehead{
\colhead{clump} & \colhead{$\Delta \chi^2$} & \colhead{$L_{H\alpha}^{\mathrm{narrow}}$} & \colhead{FWHM$^{\mathrm{narrow}}$} & \colhead{FWHM$^{\mathrm{broad}}$} & \colhead{$\Delta v$} & \colhead{F$^{\mathrm{broad}}$/F$^{\mathrm{tot}}$} & \colhead{$12+\log(O/H)$} \\
\colhead{} & \colhead{} & \colhead{($10^{41}$~erg/s)} & \colhead{(km/s)} & \colhead{(km/s)} & \colhead{(km/s)} & \colhead{} & \colhead{}} 
\startdata
$b$ & 0.17 & 1.7$\pm$0.1 &   81$\pm$6 & 270$\pm$33 &    23$\pm$11 & 0.35$\pm$0.12 & 8.02$\pm$0.08 \\
$d$ & 0.11 & 1.6$\pm$0.1 &   99$\pm$7 & 145$\pm$30 & -123$\pm$20 & 0.28$\pm$0.06 & 8.11$\pm$0.07 \\
$e$ & 0.50 & 4.0$\pm$0.2 & 109$\pm$4 & 315$\pm$10 &   -23$\pm$3   & 0.54$\pm$0.05 & 8.28$\pm$0.02 \\
$f$  & 0.12 & 0.9$\pm$0.1 &   84$\pm$8 & 270$\pm$41 &      6$\pm$9   & 0.40$\pm$0.17 & 8.08$\pm$0.06 \\
$g$ & 0.05 & 14.3$\pm$0.3 & 160$\pm$5 &                    &                       &                          & 8.07$\pm$0.06 \\ 
\enddata
\tablecomments{Columns are the improvement in the reduced $\chi^2$ between the single and double component Gaussian fits ($\sim$145 degrees of freedom); narrow-component de-lensed H$\alpha$ luminosity; FWHM of the narrow and broad component; velocity offset between both components; the ratio of broad to total flux; and oxygen abundance derived from the [N~II]/H$\alpha$ ratio. The quoted uncertainties are based on random errors,  the metallicity calibration from \cite{pp04} has a 0.2~dex systematic uncertainty and for the H$\alpha$ luminosity one should take into account an additional systematic flux uncertainty of $\sim30$\%.}
\end{deluxetable*}
\begin{figure*}
\centering
\vspace{-2.5cm}
\includegraphics[width=\textwidth]{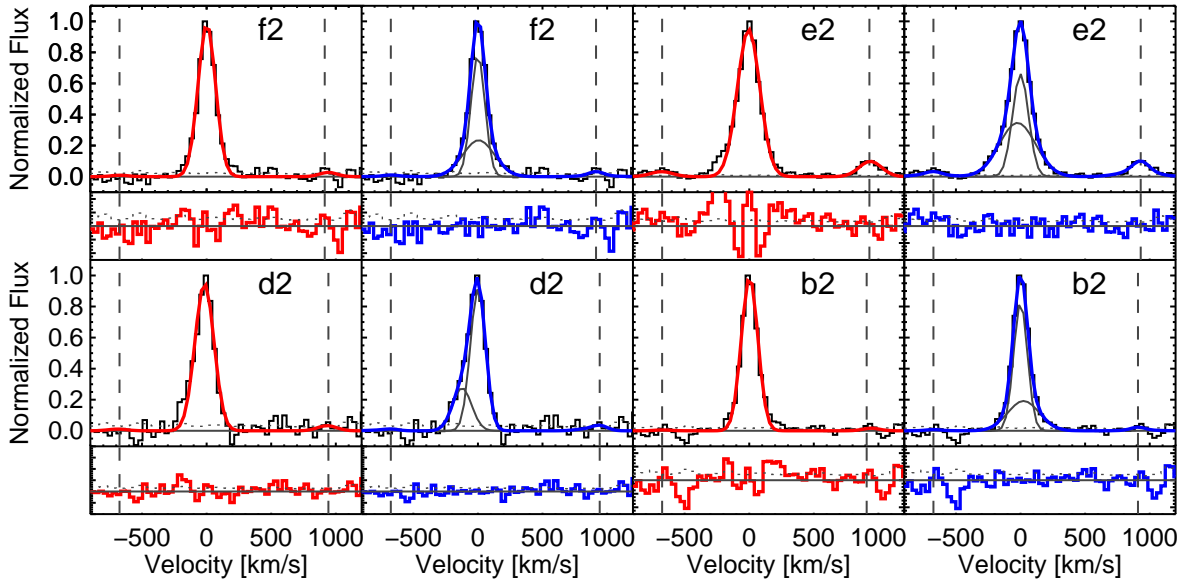}
\caption{Integrated spectra for the clumps in pointing 2, all with robust wind detections. For every clump, the best-fit single component Gaussian model is shown in red in the left panel. The right panel shows the double component model in blue, the narrow and broad component are shown separately in grey. The residuals are shown at the bottom. From these, it becomes clear that a single Gaussian component fails to fit the broad wings of the H$\alpha$ emission line, the double-component residual is significantly reduced at the location of the wings. The vertical dotted lines note the expected wavelength positions of the [N~II]~$\lambda$6548,6584 doublet. \label{fig:winds}}
\end{figure*}

\npar
In pointing 2, single Gaussian fits to the integrated spectra fail to fit the broad wings of the H$\alpha$ emission line, as can be seen in Figure~\ref{fig:winds}. The line profiles include a broad underlying component, which signifies the presence of outflows. Star formation driven galactic winds are seen in most high redshift SFGs (e.g. \citealt{Shapiro2009,Weiner2009,Rubin2010,Steidel2010}) and have recently been localized for a handful of massive, individual star-forming clumps \citep{Genzel2011,Newman2012a,Wisnioski2012}. Following \cite{Newman2012a}, we fit the H$\alpha$ line profile with a double-component Gaussian model when it improves the reduced $\chi^2$ of the fit over the $H\alpha$ and [N~II] region ($\sim$145 degrees of freedom) by at least 10\%. This is the case for all clumps in pointing 2, none of the line profiles in pointing 3 have sufficient signal-to-noise to detect an underlying broad component. The line profile parameters are reported in Table~\ref{tab:clumposiris}. The luminosities are corrected for the flux-weighted mean magnification within the OSIRIS clump aperture, and the linewidths are corrected for instrumental broadening. 
Clump star formation rates are derived from the narrow-component H$\alpha$ luminosities with the \cite{Kennicutt1998} conversion, corrected to the Chabrier IMF, and range from 0.4 to 6.6~M$_\odot$~yr$^{-1}$ (uncorrected for dust extinction). We find winds with a FWHM$^{\mathrm{broad}}=150-320$~km/s which account for 30-55\% of the total H$\alpha$ flux. These are somewhat less broad than the FWHM$^{\mathrm{broad}}\sim500$~km/s reported so far for a handful of more massive clumps in more massive $z\sim1-2$ SFGs \citep{Newman2012a,Wisnioski2012}. This is expected given the positive correlation between wind velocity/FWHM and host galaxy stellar mass \citep{Shapiro2009, Weiner2009,Newman2012b}. Both the galaxy integrated mass and clump masses of RCSGA0327 are more than an order of magnitude lower than the $z\sim2$ SFGs for which stellar winds have been resolved so far. Estimates of the mass-loading factors of outflows are highly uncertain due to the necessary assumptions on the geometry, rate and physical extent of the outflow. Following the assumptions made in \cite{Newman2012b} for a warm ionized outflow with radially constant outflow velocity, we find mass-loading factors of 1-4 times the clump SFR.

AGN feedback is a common explanation for the presence of broad emission lines. The outflows detected in RCSGA0327 are unlikely to be AGN-powered for the following reasons: 1) the diagnostic BPT diagram (as derived from the long-slit NIRSPEC and FIRE data as well as HST grism data) does not show the extreme line ratios expected for an AGN origin of the emission lines (K.~Whitaker et al. 2014, in preparation); 2) we see no evidence for point-source AGN activity in Mg~II 2800\AA\ emission or Chandra X-ray data (J.~R.~R Rigby et al. 2014, in preparation); and 3) the broad component is spatially extended, it can be identified for multiple spatial pixels within the clump apertures. Unfortunately the OSIRIS data have insufficient signal-to-noise to fit a double component Gaussian model to individual spatial pixels and spatially map the strength of the outflow. We note that the kinematic maps presented in \S\ref{subsec:kinmaps} result from single Gaussian fits. The presence of outflows will not affect the velocity map, since the velocity shift between the single best-fit model and the narrow-component of a double model is negligible. The velocity dispersion is overestimated by 30-40\% in the clump regions.

\begin{figure}
\centering
\includegraphics[width=9.5cm]{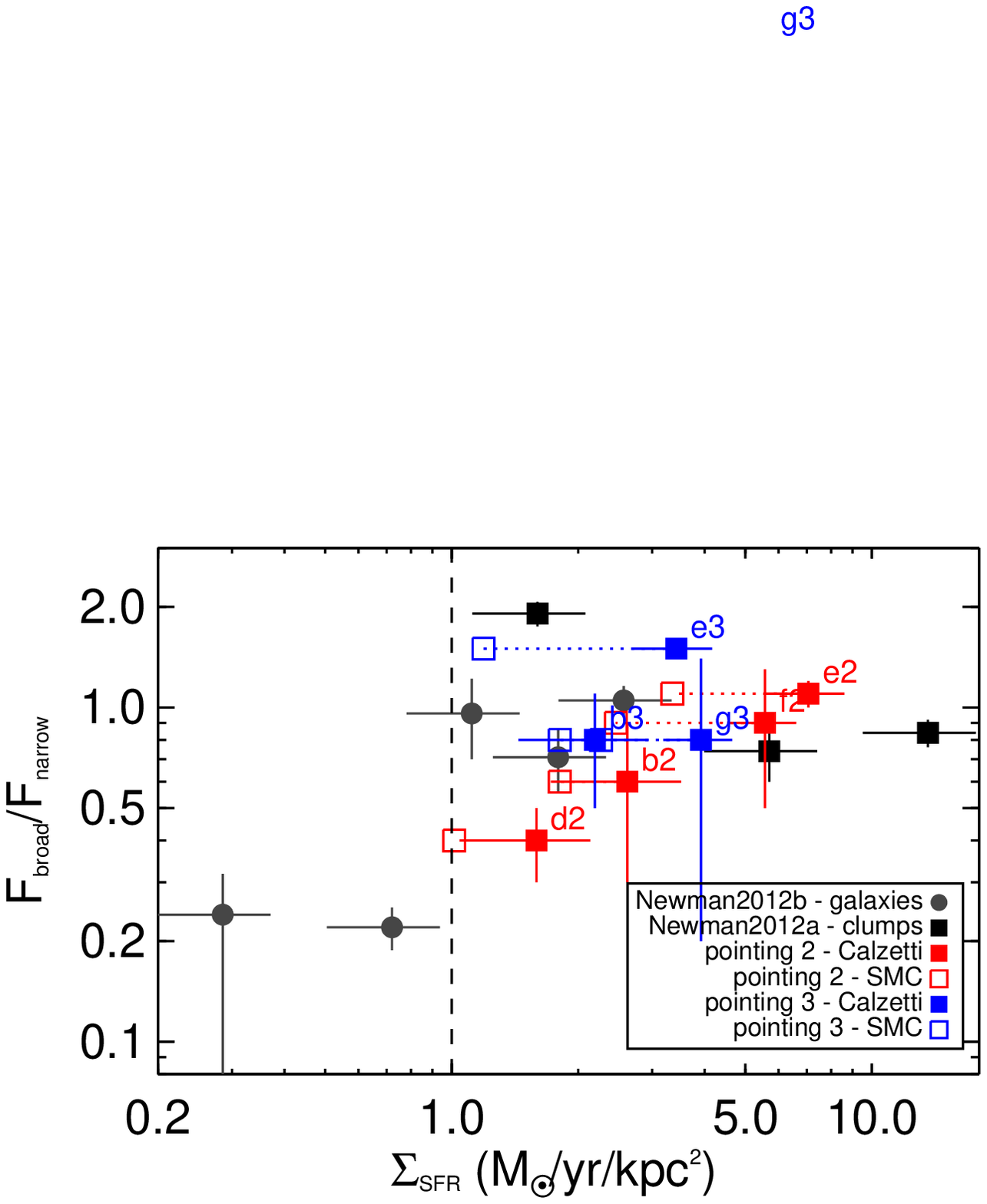}
\caption{Dependence of the outflow strength, characterized as the fraction of H$\alpha$ flux contained in the  broad versus narrow component, on star formation surface density. The grey circles present stacked results for $z\sim2$ SFGs from the SINS survey \citep{Newman2012b}. The black squares correspond to outflows detected for 3 individual clumps within two of the most massive SINS galaxies \citep{Newman2012a}. The clumps in RCSGA0327 are shown in red and blue for pointing 2 and 3 respectively. The clump SFR is derived from the narrow-component H$\alpha$ luminosity and corrected for dust extinction using the SED-derived reddening $E(B-V)_s$ from either the Calzetti (filled squares) or the SMC (open squares) dust extinction law. We assume no additional extinction towards the nebular emission lines. \label{fig:treshold}}
\end{figure}

\npar
\cite{Newman2012b} present evidence for a strong dependence of the strength of outflows on the star formation surface density ($\Sigma_{SFR}$) of the galaxy or clump from which they originate. They propose a threshold of $\Sigma_{SFR} > 1$~M$_\odot$~yr$^{-1}$~kpc$^{-2}$ to power a strong wind at $z\sim2$, i.e. where the broad component accounts for at least one third of the total flux. This is an order of magnitude higher than the SF surface density threshold found in the local Universe \citep{Heckman2002}. Since gravitational lensing conserves surface brightness, $\Sigma_{SFR}$ for the clumps in RCSGA0327 can be estimated in the image plane from the observed narrow-component H$\alpha$ luminosity and the size of the OSIRIS clump apertures. A dust correction is applied based on the SED-derived reddening, using either the Calzetti or SMC extinction law and assuming that the nebular emission lines and stellar continuum suffer the same amount of extinction (see \S\ref{subsec:clumpfirenirspec}). Figure~\ref{fig:treshold} shows that the clumps all have high SF surface densities significantly above the proposed threshold $\Sigma_{SFR} > 1$~M$_\odot$~yr$^{-1}$~kpc$^{-2}$.

\npar
The signal-to-noise of the OSIRIS observations is insufficient to detect [N~II] emission in most individual spatial pixels, but the lines are detected at S/N$>2$ in the integrated spectra of all clumps in pointing 2, as well as clumps $e$ and $g$ in pointing 3. We measure the [N~II] flux by fitting a multi-component Gaussian model to H$\alpha$ and both [N~II] lines using MPFITFUN. We fix the linewidths to a common value and constrain the ratio of the [N~II] doublet to its theoretical value of 3.071 \citep{Storey2000}. For pointing 2, a second broad component is included for all lines, also with a common linewidth. Additionally, we constrain the flux ratio of the narrow and broad component of each line to a common value. 
We estimate the metallicity of each clump from the ratio of [N~II]~$\lambda$6585 to H$\alpha$, the N2 index, as empirically calibrated by \cite{pp04}, and report the results in Table~\ref{tab:clumposiris}. The calibration of strong-line metallicity indicators remains uncertain at high redshift (e.g. \citealt{Kewley2008, me2012b}), but here we are mainly concerned with the relative variation in abundance between the clumps. The \cite{pp04} calibration has a systematic uncertainty of 0.2~dex.

\subsection{Clumps in FIRE and NIRSPEC}
\label{subsec:clumpfirenirspec}
This section explores additional information that can be learned from the long-slit near-IR spectra taken with Magellan/FIRE for clumps $u$ and $b2$ and Keck/NIRSPEC for clumps $u$ and $e2$ presented in \S\ref{subsec:data-fire} and \S\ref{subsec:data-nirspec}.

\subsubsection{Extinction}
Following R11, we measure extinctions from the NIRSPEC spectra of clumps $u$ and $e2$ from the H$\beta$/H$\gamma$ ratio, which is the brightest pair of Balmer lines covered within a single grating setting. Using the Calzetti extinction law, the measured reddening is $E(B-V)_g = 0.34\pm0.10$ for clump $u$, and $E(B-V)_g = 0.23\pm0.09$ for clump $e2$. This is a more accurate measurement than in R11, mostly because we have switched to fitting a common continuum level and linewidth for all lines in each grating setting. For FIRE, we can use both the H$\alpha$/H$\beta$ and H$\beta$/H$\gamma$ ratios. We find $E(B-V)_g^{\alpha \beta} = 0.38\pm0.04$ and $E(B-V)_g^{\beta \gamma}  = 0.32\pm0.30$ for clump $u$ and  $E(B-V)_g^{\alpha \beta}  = 0.16\pm0.03$ and $E(B-V)_g^{\beta \gamma}  = 0.07\pm0.15$ for clump $b2$. For clump $u$, the reddening estimates derived from NIRSPEC and FIRE are consistent. 

We can compare these reddening measures to the reddening of the stellar light as derived from the best-fit SED model and reported in Table~\ref{tab:clumpsed}. This comparison is visualised in the bottom left panel of Figure~\ref{fig:radial} and suggests there is no need for additional extinction towards the ionized gas when the Calzetti extinction law is applied, while $E(B-V)_g$ is significantly higher than $E(B-V)_s$ for the SMC law.

\subsubsection{Electron density}
We use the [O~II]~$\lambda$3727 doublet to constrain the electron density, using the task \textit{stsdas.analysis.temden} in IRAF\footnotemark[5] with $T_e = 10^4$~K as done in R11. From the NIRSPEC data, a line flux ratio of f(3726/3729) $= 1.19 \pm 0.07$ translates to an electron density $n_e = 600 \pm 100$~cm$^{-3}$ for clump $e2$; for clump $u$ we find f(3726/3729) $= 0.84 \pm 0.03$ and $n_e = 180 \pm 35$~cm$^{-3}$. From the FIRE data, we find f(3726/3729) $= 0.84 \pm 0.08$ and $n_e = 180 \pm 90$~cm$^{-3}$ for clump $b2$ and f(3726/3729) $= 1.01 \pm 0.13$ and $n_e =  370 \pm 150$~cm$^{-3}$ for clump $u$. 
\footnotetext[5]{IRAF is distributed by the National Optical Astronomy Observatories, which are operated by the Association of Universities for Research in Astronomy, Inc., under cooperative agreement with the National Science Foundation.}

\subsubsection{Metallicity}
We estimate the metallicity from the H$\alpha$ and [N~II] emission line fluxes measured in the NIRSPEC and FIRE spectra. Using the third-order polynomial fit of \citet{pp04}, we infer a metallicity of $12 + \log(O/H) = 8.16\pm0.02$ for clump $u$, and $12 + \log(O/H) = 8.34\pm0.02$ for clump $e2$ from the NIRSPEC data. From FIRE, we derive a metallicity of  $12 + \log(O/H) = 8.16\pm0.02$ for clump $u$ and $12 + \log(O/H) = 8.12\pm0.02$ for clump $b2$. These estimates are consistent with the OSIRIS results presented in \S\ref{subsec:clumposiris} as can also be seen in the bottom right panel of Figure~\ref{fig:radial}.

\begin{figure}
\centering
\includegraphics[width=9cm]{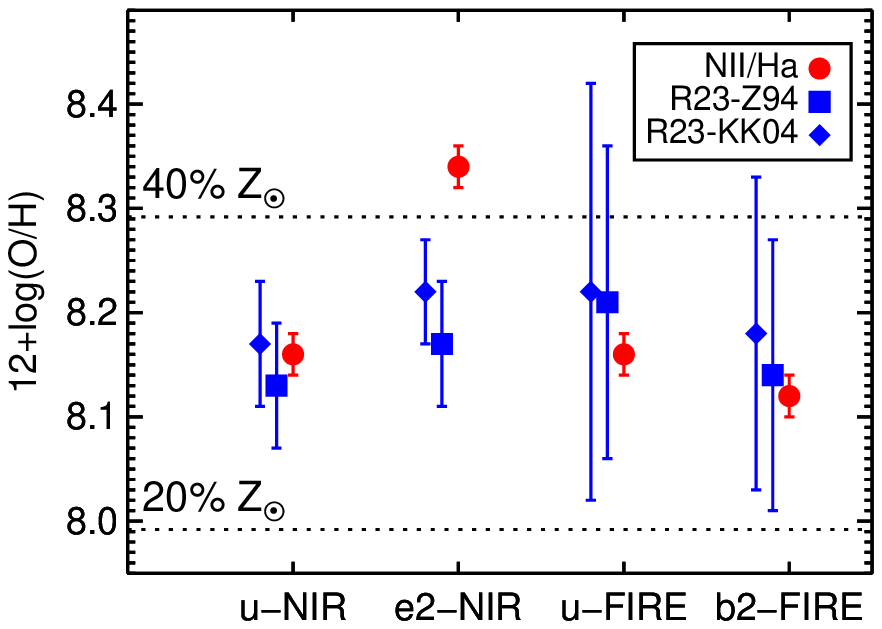}
\caption{Comparison of metallicity indicators for the NIRSPEC and FIRE spectra. We show metallicities derived from the N2 index, $\log([N~II]~\lambda6584/H\alpha)$ as calibrated by \cite{pp04} \textit{(red circles)}, as well as metallicities derived from the $R_{23}$ index, $\log{R_{23}} = \log[($[O~II]~$\lambda$3727 + [O~III]~$\lambda$4959 + [O~III]~$\lambda$5007)/H$\beta]$ as calibrated by \cite{Zaritsky1994} \textit{(blue squares)} and \cite{kk04} \textit{(blue diamonds)}. The latter two have been converted to the N2 calibration of \cite{pp04} with the conversions from \cite{Kewley2008}. Only statistical uncertainties are shown, the indicators each have a systematic uncertainty of at least 0.2~dex. \label{fig:clumpmet}}
\end{figure}

The extended wavelength coverage of the NIRSPEC and FIRE spectra includes additional lines which can be used to estimate metallicity. One needs to keep in mind the significant offsets between different strong-line indicators, which need to be converted to the same base calibration before any comparison can be made \citep{Kewley2008}. Additionally, emission line fluxes need to be corrected for dust extinction when the strong-line indicator spans a large wavelength range.
The $R_{23}$ index, $\log{R_{23}} = \log[($[O~II]~$\lambda$3727 + [O~III]~$\lambda$4959 + [O~III]~$\lambda$5007)/H$\beta]$ is commonly used in the literature. This indicator is double-valued with a low and high metallicity result for every value of $R_{23}$. Using the [N~II]/[O~II] flux ratio to distinguish between both metallicity branches \citep{Kewley2008}, we find that all clumps fall on the upper branch. We proceed to use the upper branch $R_{23}$ calibration from \cite{Zaritsky1994} as well as \cite{kk04} and convert both results to our base metallicity calibration of [N~II/H$\alpha$] from \cite{pp04}.  
The results for all clumps are shown in Figure~\ref{fig:clumpmet}. Only statistical uncertainties are shown, the indicators each have a systematic uncertainty of at least 0.2~dex. The larger uncertainties for the $R_{23}$ indicators originate from the propagation of the uncertainty in the reddening. We see a general agreement between metallicity indicators, except for clump $e2$, where the metallicity derived from the [N~II]/H$\alpha$ ratio is larger by $\sim0.15$~dex. This offset falls within the significant systematic uncertainties involved in the comparison of metallicity indicators, but given the agreement between the N2 and $R_{23}$ index for the other clumps, it does suggest that the [N~II]/H$\alpha$ ratio of clump $e2$ is elevated. This agrees with the higher electron density measured for this clump, which results in an increased rate of collisional excitation.
Additionally, the galactic wind detected for clump $e$ could drive shock excitation. At high redshift ($z>1.5$), the presence of slow shocks mimics a higher metallicity starburst \citep{Kewley2013}.

\subsubsection{Outflows}
The FIRE spectrum of clump $b2$ and the NIRSPEC spectra of clump $e2$ show an outflow consistent with the OSIRIS results. No outflow is detected in the FIRE or NIRSPEC spectra of clump $u$.

\subsection{Radial Variation of Clump Properties}
\label{subsec:radial}
\begin{figure*}
\centering
\includegraphics[width=\textwidth]{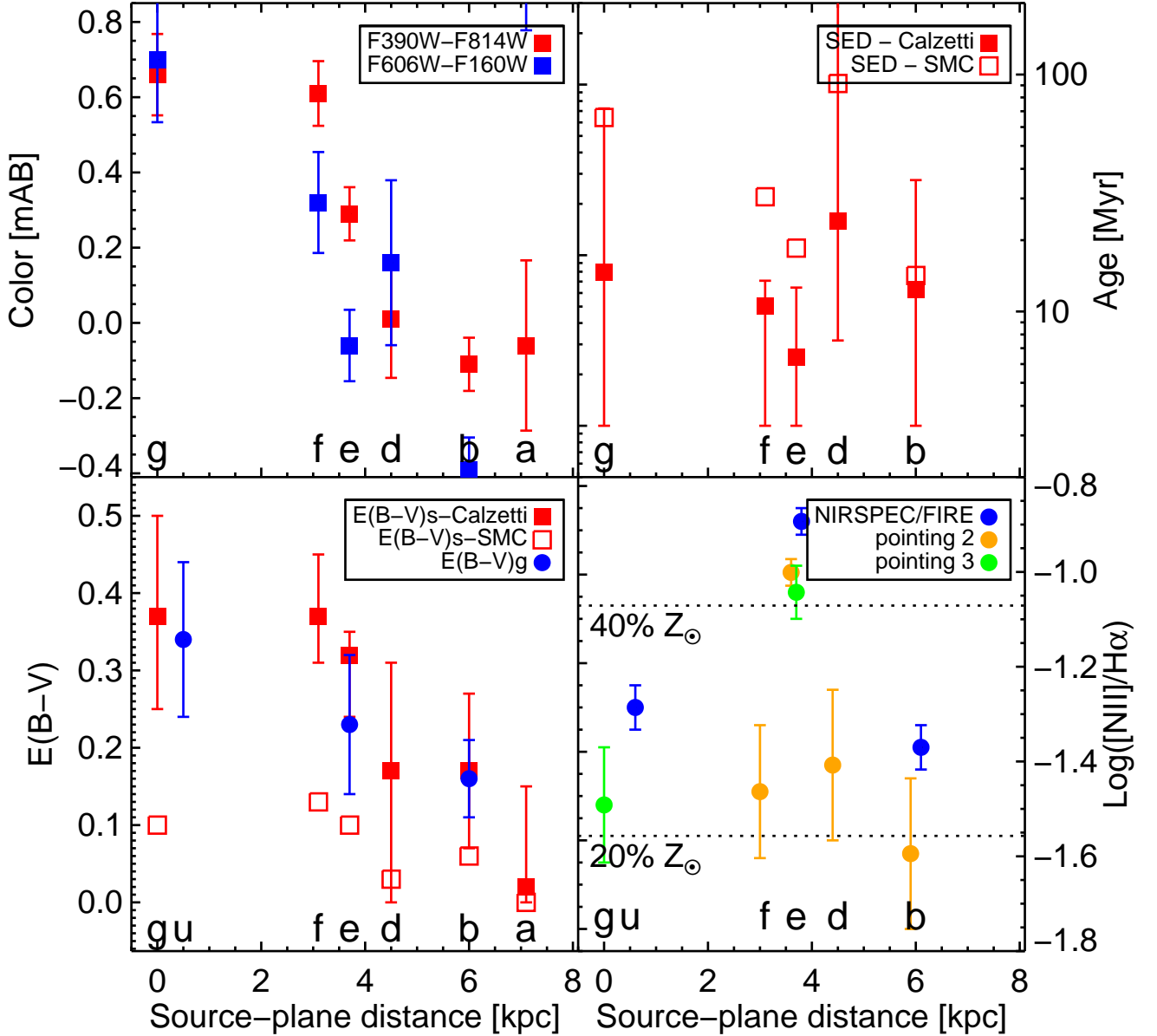}
\caption{Radial variation of clump properties. The x-axis corresponds to the projected source-plane distance from each clump to clump $g$. Red and blue symbols denote results from the SED fit with the Calzetti and SMC extinction laws respectively. Orange and green symbols correspond to OSIRIS results, blue symbols show long-slit spectroscopy with NIRSPEC and FIRE. Errorbars only reflect statistical uncertainties. \textit{(Top left)} F390W-F814W rest-frame UV color; \textit{(Top right)} stellar age; \textit{(Bottom left)} reddening from the SED fit and from the Balmer ratio; \textit{(Bottom right)} metallicity. \label{fig:radial}}
\end{figure*}
Any trends in clump properties with respect to their location within the galaxy can provide additional constraints on the clump origin. No correlation is expected between the properties of separate interacting components. Since there is no clear definition of the galaxy center for RCSGA0327, we adopt the position of clump $g$, which is the brightest clump and corresponds to a strong peak in stellar mass surface density (see \S\ref{sec:spatialsed}).
For each clump, the projected distance to clump $g$ is measured in the reconstructed source-plane images. 
The top left panel of Figure~\ref{fig:radial} shows the rest-frame $U-V$ color F606W-F160W as well as a measure of the UV-slope from the F390W-F814W color. We see a slight radial trend, where clumps become redder by 0.5-1~mag when moving East from clump $a$ towards clump $g$. Such color trends have been interpreted as evidence for a radial age trend, confirming a picture of radial migration of clumps formed through gravitational collapse of a turbulent disk \citep{Guo2012}. However, rest-frame UV color is governed by a degeneracy between age, metallicity and dust extinction, and a color trend can be caused by a trend in any of these parameters. The top right panel shows the stellar age as derived from the SED using both the Calzetti and SMC extinction laws (filled and open red squares respectively).
The stellar age remains roughly constant across the clumps, within the significant uncertainties. There is some evidence for an increase in reddening towards clump $g$, as shown in the bottom left panel. Some correlation between color and reddening is expected from the SED modelling, but the trend is confirmed by the reddening of the ionized gas derived from the NIRSPEC and FIRE spectra (blue filled circles). 

The bottom right panel of Figure~\ref{fig:radial} shows the [N~II]/H$\alpha$ ratio, as measured in OSIRIS as well as the NIRSPEC and FIRE spectra. The [N~II]/H$\alpha$ ratio shows a mostly flat gradient across the clumps, with the exception of clump $e$. We have discussed in \S\ref{subsec:clumpfirenirspec} how the elevated [N~II]/H$\alpha$ ratio for this clump is likely due to an increased electron density and/or shock excitation in the outflow. Based on the $R_{23}$ indicator, the metallicity of clump $e$ agrees with the other clumps. A flat metallicity gradient supports the scenario of an ongoing interaction within the system. In merger simulations, galaxy metallicity gradients are found to flatten as the merger progresses and low-metallicity gas is transported from the outskirts of the interacting galaxies to the central region \citep{Rupke2010a}. This has been seen in local close-pair spiral galaxies \citep{Kewley2010, Rupke2010b} and local LIRGs \citep{Rich2012}. Integral-field spectroscopy studies of both lensed and non-lensed $z\sim2$ SFGs are also finding flatter or even inverted metallicity gradients in interacting systems, though samples are still small \citep{Jones2013}.

\subsection{Clump Scaling Relations}
\label{subsec:scaling}
\begin{figure*}
\centering
\includegraphics[width=\textwidth]{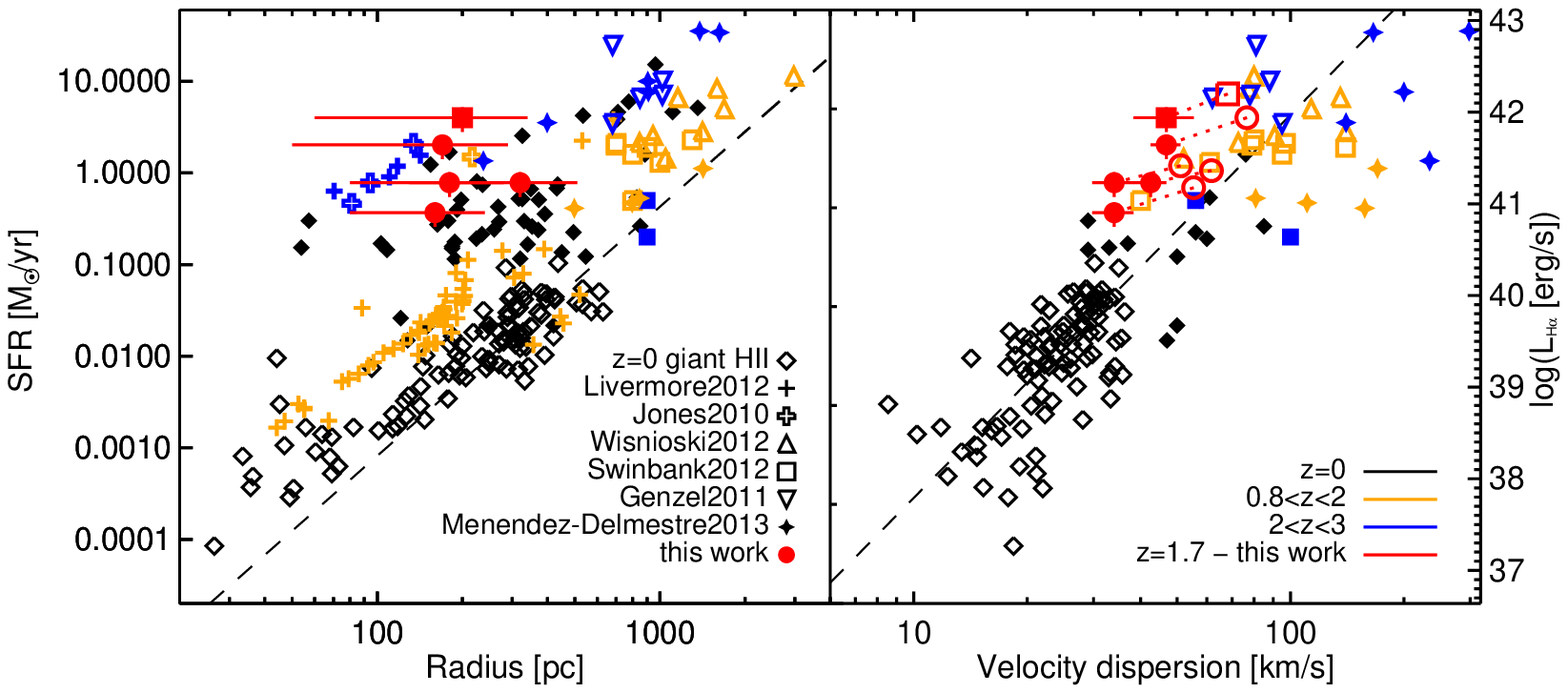}
\caption{Scaling relations between H$\alpha$ luminosity and size \textit{(left)} or velocity dispersion \textit{(right)} for local and high redshift star-forming clumps. The dashed black lines display the best-fit scaling relations from \cite{Wisnioski2012}. The local sample consists of giant H~II regions in local spirals (rotating disks; \citealt{Gallagher1983, Arsenault1988,Rozas2006}); and giant H~II regions in local ULIRGs (interacting systems; \citealt{Bastian2006,Monreal2007,Rodriguez2011}). We use open and closed black diamonds to differentiate between these kinematic classifications. The high-z clumps come from six different studies mentioned in the text and legend. We have color-coded them by redshift to look for redshift evolution of the scaling relations: $0.8<z<2$ (orange) and $2<z<3$ (blue). Open and closed symbols are again used to differentiate between kinematically classified rotating disks and interacting systems. The lensed galaxies from \cite{Livermore2012} do not have kinematic information and are shown with plus-symbols. Neither \cite{Livermore2012} nor \cite{Jones2010} report velocity dispersion measurements for their clumps, which are therefore not included in the right panel. Results from the narrow-component H$\alpha$ emission line profile for the clumps in RCSGA0327 are shown with filled red circles and a filled red square for clump $g$ . The open red symbols in the right panel show the overestimate of both H$\alpha$ luminosity and linewidth when fitting the line profiles with a single Gaussian, not taking into account the broad underlying wind component. \label{fig:scaling}}
\end{figure*}

The five well-measured clumps in RCSGA0327 present a sizeable contribution to the current sample of 40 star-forming regions in $z\sim2$ SFGs with reliable H$\alpha$ measurements: a) five clumps from three $z\sim2$ SINS galaxies \citep{Genzel2011}; b) eight clumps from three massive SFGs at $z\sim1.3$ from the WiggleZ survey \citep{Wisnioski2012}; c) eight clumps from four lensed galaxies at $z=1.6-2.6$ \citep{Jones2010}; d) nine clumps from four H$\alpha$-selected galaxies at $z=1.4$ and $z=2.2$ from HiZELS \citep{Swinbank2012}; and e) ten clumps in three submillimeter-selected galaxies (SMG) at $z=1.4-2.4$ \citep{Menendez2013}. Additionally, \cite{Livermore2012} use HST/WFC3 narrowband imaging centered on H$\alpha$ to study clump sizes and luminosities in an additional eight lensed galaxies at $z=1-1.5$; they have no kinematic information. 
Figure~\ref{fig:scaling} compares H$\alpha$ luminosity (uncorrected for dust extinction), size and velocity dispersion measurements for the sample of high-z clumps to local scaling relations between these parameters taken from \cite{Wisnioski2012}. These authors have remeasured all clump sizes consistently with 2D elliptical Gaussian fits. The size-luminosity relation in the left panel clearly shows how the three lensing studies (\citealt{Jones2010, Livermore2012} and this work) probe clump sizes up to an order of magnitude smaller than what can be resolved in non-lensed studies. The clumps in RCSGA0327 are broadly consistent with the other high-z clumps and lie roughly two orders of magnitude above the local luminosity-size scaling relation. As was first pointed out by \cite{Livermore2012}, the offset seems to increase with redshift. These authors found correlations between clump SFR surface density and the SFR surface densities and gas surface densities of the host galaxies. As such, high redshift clumps appear to be scaled-up analogues of local H~II regions, simply bigger and brighter because of the increasing gas fractions in high-z SFGs.

It is worth asking whether the dynamical state of the host galaxy plays a role in determining the clump SFR surface densities. In the local Universe, giant H~II regions found in interacting systems (shown with black filled diamonds) show systematically higher SFR surface densities compared to giant H~II regions in local spirals (open diamonds). At high redshift, we do not see elevated SFR surface densities for clumps within kinematically classified interacting systems (RCSGA0327, two lensed galaxies from \cite{Jones2010}, one HiZELS source from \cite{Swinbank2012} and three SMGs studied by \cite{Menendez2013}). It would be very valuable to obtain direct gas measurements of these systems to clarify the connection between gas surface density, SF surface density and galaxy kinematics.

\npar
\begin{figure*}
\centering
\includegraphics[width=\textwidth]{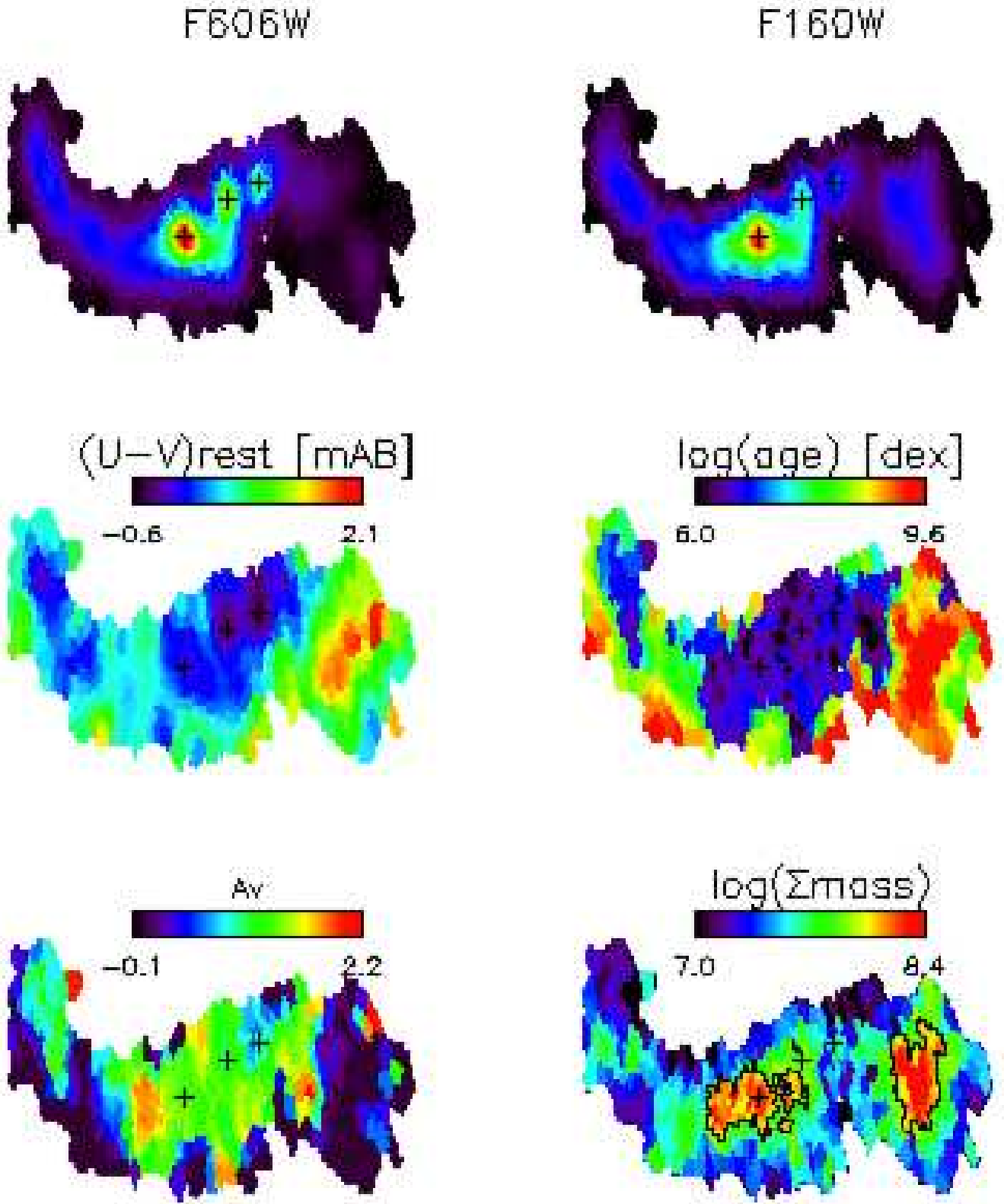}
\caption{Source-plane images of RCSGA0327 based on the counter-image: surface brightness distributions in the F606W and F160W bands (corresponding to rest-frame 2800\AA\ and 5500\AA); rest-frame U-V color map based on F814W-F160W; stellar age; dust extinction; and stellar mass surface density. Clumps $g$, $e$ and $b$ are marked by black crosses, from left to right. The contour on the stellar mass surface density map corresponds to $\log(\Sigma_{M*})=8.0$. \label{fig:resolvedsed}}
\end{figure*}

A few caveats should be kept in mind in the analysis of the scaling relations presented above. First, the high redshift clumps form by no means a uniformly selected sample, but span a large range of host galaxy selection and integrated properties. Secondly, the H$\alpha$ luminosities in Figure~\ref{fig:scaling} have not been corrected for dust extinction. This is mostly driven by the lack of reliable extinction estimates, especially on the scale of the individual clumps. Variations in dust extinction with redshift or host galaxy properties could have a significant effect on the scatter in these scaling relations. Finally, not taking into account a broad underlying wind component when fitting the H$\alpha$ emission line profile will lead to overestimates of both the H$\alpha$ luminosity and velocity dispersion of clumps. This is illustrated for RCSGA0327 with the open and closed red symbols in the right panel of Figure~\ref{fig:scaling}.

\section{Spatially Resolved SED Modelling}
\label{sec:spatialsed}

\begin{figure*}
\centering
\includegraphics[width=\textwidth]{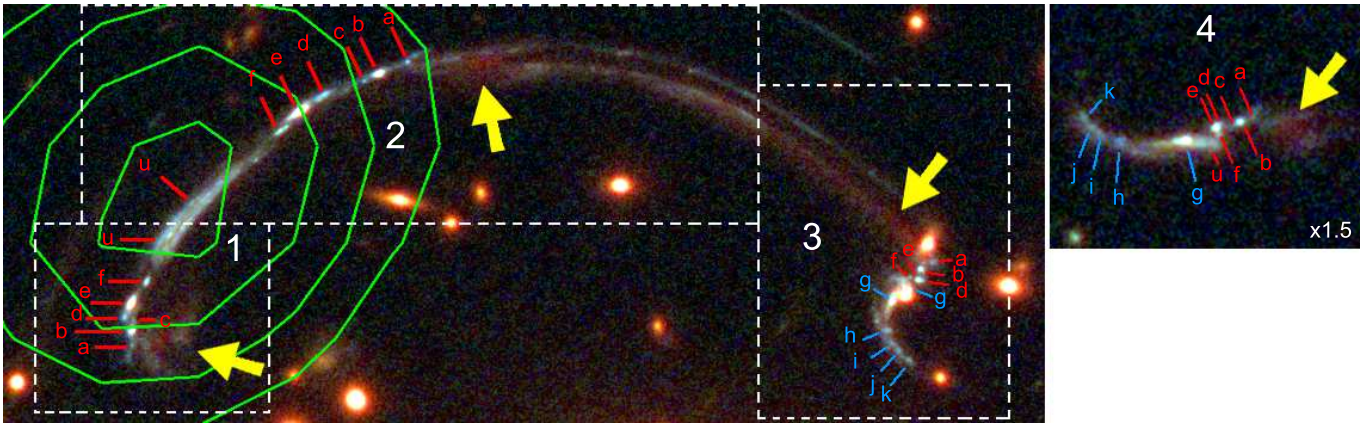}
\caption{Same as Figure~\ref{fig:implane}. The green lines correspond to the 1.5-2.0-2.5-3.0$\sigma$ contours of the PACS 100~$\micron$ image, which represents the best combination of signal-to-noise, spatial resolution, and sensitivity to re-processed dust emission out of the available far-infrared imaging. The yellow arrows indicate the second, red mass component in each of the four images of the source. The Herschel data make clear that the bulk of the long wavelength emission is associated with clump $u$ and that there is no emission associated with this second mass concentration. \label{fig:herschel}}
\end{figure*}

The stellar mass surface density of a galaxy holds crucial information regarding the physical origin of its individual star-forming regions. When clumps correspond to separate interacting components, one can expect a well-established older stellar population underlying their strong rest-frame UV presence caused by the recent star formation triggered in the interaction. In contrast, clumps formed through gravitational collapse of a gas-rich, turbulent disk are too short-lived to build up a significant population of old stars. Dynamical friction against the underlying galaxy disk and clump-clump interactions cause the clumps to spiral inwards and coalesce into the galaxy center on timescales $\sim300-500$~Myr (e.g. \citealt{Dekel2013}). The detection of strong outflows originating from clumps in RCSGA0327 and other studies \citep{Genzel2011,Newman2012a,Wisnioski2012} could disrupt the clumps on even shorter timescales \citep{Genel2012}.
Thus, while clumps dominate the galaxy morphology in rest-frame UV light which mostly traces newly-formed O and B stars, they become much less prominent at rest-frame optical wavelengths. Along these lines, \cite{Wuyts2012} have recently quantified the reduced contribution of clumps to stellar mass maps of clumpy galaxies at $0.5<z<2.5$ in the CANDELS fields based on spatially resolved SED modelling. Here we perform similar modelling for RCSGA0327.
\npar
We model the counter-image to obtain a full image of the source-plane galaxy without the contamination of the cluster galaxies that fall on top of image 3. Following the procedure outlined in \cite{Wuyts2012}, we PSF-match the different WFC3 images to the broadest F160W PSF and group pixels using 2D Voronoi binning \citep{Cappellari2003} to achieve S/N$\ge$10 in this band. For the SED fit, we use the default assumptions described in \S\ref{subsec:clumpsed} for the modelling of the clumps: BC03 models, Calzetti dust extinction, Chabrier IMF, 0.2-0.4~Z$_\odot$ metallicity, exponentially decreasing SFHs with $\log(\tau)\ge8.5$, no age restriction. Different modelling assumptions will generally affect only the absolute value of the derived stellar population parameters and here we are mostly interested in the relative variation of these parameters across the galaxy. As discussed in \S\ref{subsec:clumpsed}, the F098M and F125W filters can be heavily contaminated by line emission and are excluded from the SED fit. As a consistency check, adding the stellar mass of all the Voronoi bins returns a total galaxy mass within 0.1~dex of the stellar mass derived from the galaxy-integrated photometry; \cite{Wuyts2012} found a scatter of 0.08~dex in their comparison of integrated and resolved stellar masses for $1.5<z<2.5$ CANDELS galaxies.
\npar
The top row of Figure~\ref{fig:resolvedsed} shows source-plane maps of the F606W and F160W WFC3 images, which roughly correspond to rest-frame 2800\AA\ and 5500\AA. Clumps $g$, $e$ and $b$ are shown with black crosses, from left to right. The clumps become less pronounced at redder wavelengths, but clump $g$ still shows up strongly in the F160W image. This agrees with its significant presence in the stellar mass surface density map in the bottom right panel of Figure~\ref{fig:resolvedsed}. Based on this map, the other large mass concentration in RCSGA0327 lies on its western edge, to the right of all the clumps. This region has not been discussed so far, since it was not detected in H$\alpha$ emission\footnotemark[6]. It is very red in the false-color image presented in Figure~\ref{fig:implane} and the rest-frame $U-V$ color map in Figure~\ref{fig:resolvedsed}. Based on the stellar age and dust extinction maps, the red color is attributed to an old stellar population, and not to dust-obscured star formation. This agrees with the non-detection of this region in the OSIRIS data and recent Herschel PACS/SPIRE observations (Figure~\ref{fig:herschel}; J.~R.~Rigby et al.~2014, in preparation). 

Out to an isophote of $\log(\Sigma_{M*}) \ge 8.0$ (the black contour on top of the stellar mass surface density map in Figure~\ref{fig:resolvedsed}), clump $g$ and the red western region each contain $\sim1.2 \times 10^9$~M$_\odot$, or $\sim20$\% of the galaxy-integrated stellar mass in RCSGA0327. Within this contour, the western mass component has an intrinsic SFR of 0.5~M$_\odot$/yr, which puts it $\sim2\sigma$ below the main sequence at $z=2$.
\footnotetext[6]{Due to the non-detection in the OSIRIS data, we have no spectroscopic confirmation of the redshift of this red component. However, given that it is multiply imaged together with the rest of the system, the details of the lens modelling restrict its redshift to within $\sim1000$~km/s of systemic velocity.}

\section{Summary of observational results and discussion}
\label{sec:disc}
We have presented a detailed analysis of the kinematics, spatially resolved stellar population parameters and clump properties of RCSGA0327 based on multi-wavelength HST/WFC3 imaging and AO-assisted OSIRIS IFS data. The main results are summarized below.
\begin{itemize}
\item The kinematical analysis of the OSIRIS data strongly suggests an ongoing interaction, which has caused a large tidal tail extending from clump $g$ towards the North-East. Velocity dispersion peaks between the clumps could arise from high turbulence due to the interaction or overlapping H$\alpha$ emission along the line of sight. 
\item We have identified seven individual star-forming regions in the WFC3 imaging with diameters ranging from 300 to 600~pc. SED modelling of the clump photometry predicts stellar masses $10^7$ - $5 \times 10^8$~M$_\odot$, young ages of $\sim5-100$~Myr and low reddening $E(B-V)<0.4$. The steeper SMC extinction law is preferred over the default Calzetti law to avoid unphysical high clump SFRs compared to the H$\alpha$-derived values as well as the galaxy-integrated SFR.
\item RCSGA0327 is the lowest mass high-z SFG to date with resolved outflows. We find broad underlying wind components in the H$\alpha$ emission line profile of four clumps, contributing on average $\sim40$\% to their total H$\alpha$ flux. The SFR surface densities of these clumps all fall above the high-z threshold of $\Sigma_{SFR} > 1$~M$_\odot$~yr$^{-1}$~kpc$^{-2}$ to power strong winds. 
\item We find a radial gradient in rest-frame UV color of the clumps across the galaxy, which we infer to be caused by a gradient in reddening. The stellar age of the clumps, though uncertain in absolute value, remains roughly constant. We find a flat metallicity gradient, as expected for an interacting system.
\item The clumps in RCSGA0327 agree with the size-luminosity and dispersion-luminosity correlations inferred from earlier lensed and non-lensed studies of high-z clumps. Kinematical classification of the host galaxy as an interacting system does not result in higher clump SFR surface densities, unlike what has been seen for local giant H~II regions.
\item Stellar mass surface density maps based on spatially resolved SED modelling suggest an established stellar population at the location of clump $g$ and a second mass component at the western edge of the galaxy. Both contain $\sim20$\% of the total stellar mass of the system. The western mass component is not detected in H$\alpha$ or far-IR emission.
\end{itemize}
\npar
In contrast to most clumpy $z\sim2$ SFGs that have been studied in detail so far, RCSGA0327 does not agree with a single turbulent rotating disk where clumps have formed through gravitational collapse. The system is undergoing an interaction, which has boosted the specific SFR to a factor 3 above the main-sequence at $z\sim2$. The enhanced star formation is localised in multiple compact star-forming regions with high star formation surface densities and signatures of outflows. Such off-center areas of enhanced SF activity have also been seen in local mergers, such as the Antennae galaxies \citep{Whitmore1995, Karl2010}. The red mass component at the western edge of the system does not show significant dust-obscured or unobscured star formation. As such, this component must have been gas-poor before the start of the interaction, such that no new burst of SF could be triggered. This would also explain why we don't see a tidal tail originating from this component. We note that for clump $g$, the young stellar age derived from the SED fit is only relevant for the current SF episode triggered by the interaction, and does not contradict the interpretation of this clump as an established stellar population.

As a roughly equal-mass, mixed merger of one gas-rich and one gas-poor component, RCSGA0327 is not a common occurrence. Both theoretical and observational estimates of merger rates find that only $\sim10$\% of $z\sim2$ galaxies at $10^9$~M$_\odot$ are currently undergoing an interaction \citep{Conselice2003,Guo2008}. However, the number of mergers involving one or both components to be gas-poor is significantly less than that, given the overall increase in galaxy gas fraction with redshift. \cite{Lin2008} find that 24\% of mergers are mixed at $z\sim1.1$ based on galaxy pair counts in the DEEP2 Redshift Survey. On top of that, the stellar mass estimate of $\sim 1.2 \times 10^9$~M$_\odot$ for the gas-poor component in RCSGA0327 is unusually low, as can be seen for example from the mass distribution of passive galaxies at $1.4 < z < 2.5$ in GOODS-South (Figure 3, \citealt{Lee2013}).

Merger simulations mostly focus on the low-redshift Universe and have so far failed to take into account the different nature of $z\sim2$ SFGs as evidenced by their clumpy morphology, higher gas fractions and stronger turbulence. Recent studies have shown that turbulence and clumpiness have a substantial effect in mergers of present-day spirals with just a few percent of gas, causing significant differences in the star formation history during the interaction \citep{Teyssier2010, Saitoh2009}. The effect could presumably be more dramatic in high-redshift mergers involving high gas fractions. \cite{Bournaud2011} present the first wet merger simulations involving two realistic massive, gas-rich, clumpy disks. 
Such work needs to be extended to lower mass galaxies, as well as mixed and dry mergers involving early type, gas-poor components.
\npar
To finish, we would like to stress two main points which were instrumental in obtaining a full understanding of the physical nature of RCSGA0327. The first point concerns the combination of IFS data with high-resolution rest-frame UV to optical imaging. The kinematics of the ionized gas and the morphology of current star formation derived from IFS data need to be complemented with an understanding of the underlying stellar population derived from spatially resolved SED modelling. In the case of RCSGA0327, the gas-poor component in the interaction only became apparent in the stellar mass surface density maps. Secondly, the analysis presented here would not have been possible without the lensing magnification, which allowed a high-resolution velocity profile and detailed measurements of multiple $<1$~kpc size clumps. The resulting unprecedented view of a rare ongoing interaction at $z\sim2$ shows the promise of detailed study of individual systems to aid and constrain theoretical efforts towards understanding galaxy formation and evolution.

\vspace{0.5cm}
\begin{acknowledgments}
We thank our anonymous referee for a thorough reading of the paper and insightful comments. E.~W. thanks John Hibbard, Tucker Jones, Rachael Livermore, Chris Mihos, Thorsten Naab, Sarah Newman, Emily Wisnioski and Tian-Tian Yuan for sharing data and/or stimulating discussions. 
Support for HST program 12267 was provided by NASA through a grant from the Space Telescope Science Institute, which is operated by the Association of Universities for Research in Astronomy, Inc., under NASA contract NAS 5-26555.
Travel support for the Keck observations was provided by the Grants-in-Aid of Research Program of the Sigma Xi Scientific Research Society and the NASA Keck PI Data Award, administered by the NASA Exoplanet Science Institute. 
K.S acknowledges support from the University of Michigan's Presidential Fellowship. 

Data presented in this paper were partly obtained at the W.M. Keck Observatory from telescope time allocated to the National Aeronautics and Space Administration through the scientific partnership with the California Institute of Technology and the University of California. The Observatory was made possible by the generous financial support of the W.M. Keck Foundation. We acknowledge the very significant cultural role and reverence that the summit of Mauna Kea has always had within the indigenous Hawaiian community. We are most fortunate to have the opportunity to conduct observations from this mountain.
\end{acknowledgments}


\end{document}